\documentclass[]{interact}
\usepackage{bm}
\usepackage{amsfonts, amsmath, amssymb}
\usepackage{epstopdf}
\usepackage[caption=false]{subfig}
\usepackage{xcolor} 
\usepackage[colorlinks = true,
            linkcolor = blue,
            urlcolor  = blue,
            citecolor = blue,
            anchorcolor = blue]{hyperref}

\usepackage[numbers,sort&compress]{natbib}
\bibpunct[, ]{[}{]}{,}{n}{,}{,}
\renewcommand\bibfont{\fontsize{10}{12}\selectfont}

\theoremstyle{plain}

\theoremstyle{definition}

\theoremstyle{remark}

\usepackage{subfig}
\usepackage{xurl}
\DeclareMathOperator*{\argmax}{arg\,max}
\DeclareMathOperator*{\argmin}{arg\,min}
\allowdisplaybreaks
\usepackage{lineno}

\begin{document}
\articletype{}

\title{Estimating Changepoints in Extremal Dependence, Applied to Aviation Stock Prices During COVID-19 Pandemic}

\author{
\name{Arnab Hazra and Shiladitya Bose \thanks{Contact: Arnab Hazra (Email: \href{mailto:ahazra@iitk.ac.in}{ahazra@iitk.ac.in})}}
\affil{Department of Mathematics and Statistics, Indian Institute of Technology Kanpur, Kanpur, India 208016.}
}
\maketitle


\begin{center}
\begin{abstract}
The dependence in the tails of the joint distribution of two random variables is generally assessed using $\chi$-measure, the limiting conditional probability of one variable being extremely high given the other variable is also extremely high. This work is motivated by the structural changes in $\chi$-measure between the daily rate of return (RoR) of the two Indian airlines, IndiGo and SpiceJet, during the COVID-19 pandemic. We model the daily maximum and minimum RoR vectors (potentially transformed) using the bivariate H\"usler-Reiss (BHR) distribution. To estimate the changepoint in the $\chi$-measure of the BHR distribution, we explore two changepoint detection procedures based on the Likelihood Ratio Test (LRT) and Modified Information Criterion (MIC). We obtain critical values and power curves of the LRT and MIC test statistics for low through high values of $\chi$-measure. We also explore the consistency of the estimators of the changepoint based on LRT and MIC numerically. In our data application, for RoR maxima and minima, the most prominent changepoints detected by LRT and MIC are close to the announcement of the first phases of lockdown and unlock, respectively, which are realistic; thus, our study would be beneficial for portfolio optimization in the case of future pandemic situations.
\end{abstract}   

\end{center}

\section{Introduction}
\label{sec:Intro}

The COVID-19 pandemic had a profound and disruptive impact on a global scale, causing not only a substantial loss of human lives but also significant economic turmoil. Nearly every industry faced adverse consequences as a result, including the aviation sector. The outbreak of the pandemic triggered a severe decline in air travel demand, as countries implemented travel restrictions and lockdown measures to curb the spread of the virus. Extensive research, such as the study conducted by the International Civil Aviation Organization \citep{icao2020effects} on the effects of COVID-19 on worldwide Civil Aviation, has examined these repercussions. Indian airlines encountered numerous challenges during this unprecedented period, as highlighted in \cite{kaur2021impact} and \cite{agrawal2021sustainability}, who examined the specific difficulties faced by Indian airlines due to the pandemic. The imposition of domestic and international flight suspensions for several months by the Government of India had a significant financial impact on airlines, leading to layoffs, salary cuts, and even bankruptcies \citep{toi2020indigo}.

The two airlines IndiGo and SpiceJet have played crucial roles in fostering the growth of the Indian aviation sector. These airlines have effectively increased the accessibility and affordability of air travel for a broader range of people. Operating in a highly competitive market, they have continuously expanded their flight routes, enhanced services, and taken innovative features to attract passengers \citep{banerji2016case}. Thus, it is unavoidable that there exists an interdependence within their businesses. Analyzing the share prices of the two companies over a period of time allows one to gauge the market's perception of the companies and their financial performances. There are several parameters to consider while using share prices for performance evaluation, such as historical share price trends, relative performance, market index comparisons, and the impact of dividends and stock splits. The traders emphasize the importance of \textit{buying low and selling high} while aiming to generate profits. This principle underscores the significance of timing and strategic decision-making in the context of stock market investments \citep{cartea2014buy}. During the first phase of the lockdown, India and the rest of the world witnessed a significant decline in the stock market for many shares, primarily caused by reduced investments; at the same time, once the lockdown period was over, the stock prices started regaining \citep{gormsen2020coronavirus}. While analyzing daily average stock prices gives an understanding of the bulk of the distribution of the share prices, our goal is to focus on understanding the upper and lower tail behaviors of stock prices, which are more precise indicators of their interdependence at the sell and buy positions, respectively. Various companies employed diverse strategies to mitigate substantial losses during the pandemic \citep{bt2020indigo}. As a result, the tail interdependence, i.e., the tendency of selling both SpiceJet and IndiGo stocks simultaneously in high amounts or buying both stocks simultaneously in low amounts, could have experienced significant change during the COVID-19 pandemic. To analyze them effectively, we employ joint modeling in the context of extreme value theory (EVT) and changepoint analysis.

 In the context of joint modeling, the dependence structure between two random variables is often described using bivariate copulas. They are mathematical functions providing a way to model the joint distribution of variables while separately modeling their marginal distributions \citep{nelsen2006introduction}. Copulas offer flexibility in capturing various forms of dependence, including linear, nonlinear, and tail dependence, making them valuable in finance, insurance, and risk management \citep{cherubini2004copula, jaworski2010copula, patton2012review}. While the Gaussian copula is the most commonly used copula, it is criticized in the context of analyzing tail-interdependence because of its thin joint tails \citep{davison2013geostatistics}. Several alternatives like Student-$t$ copula and other elliptical models have been explored in the literature \citep{hashorva2010residual}. In the context of investigating temporally-varying tail dependnece between several crypto-currencies like Bitcoin and Ethereum, \cite{gong2022asymmetric} propose a new flexible copula approach that allow both dependence as well as independence in the tails through varying the model parameters.

In recent years, EVT has emerged as one of the most crucial fields of study in different scientific disciplines \citep{coles2001introduction, davison2015statistics}. The main applications of EVT are in areas such as portfolio management in actuarial studies \citep{huang2020lifetable}, financial risk assessment \citep{valerie2016extreme}, telecom traffic prediction \citep{molina2008application}, and detection of meteorological changes \citep{zwiers1998changes}. In financial extremes, \cite{rocco2014extreme} showed that a portfolio is more affected by a few extreme movements in the market than by the sum of many small-scale shifts. The majority of EVT focuses on methods for analyzing univariate extremes; however, \cite{tawn1988bivariate} and \cite{oliveira1989statistical} introduced the statistical methodology for bivariate extremes and \cite{coles1994statistical} further studied multivariate extremes. While the Pearson correlation coefficient measures dependence in the bulk of the joint distribution of two random variables, $\chi$-measure \citep{sibuya1960bivariate} assesses the dependence in tails. A bivariate Gaussian distribution is the most common model for bivariate responses; however, its components are asymptotically independent for any non-trivial choice of the Pearson correlation coefficient \citep{sibuya1960bivariate}. In this context, \cite{Husler1989maxima} obtained the limiting distribution of block maxima or block minima of bivariate Gaussian distribution under certain assumptions on the correlation between the components, and their bivariate model is known as the bivariate H\"usler-Reiss (BHR) distribution. The BHR distribution and its infinite-dimensional case, called Brown-Resnick process \citep{brown1977extreme}, have been used in numerous studies in the context of bivariate, multivariate, as well as spatial extreme value analysis \citep{kabluchko2009stationary, huser2013composite}. Besides, the BHR distribution is a building block of the high-dimensional graphical models for extremes \citep{engelke2020graphical, engelke2021sparse}.

A changepoint is a place or time where the statistical properties of a sequence of observations change; in other words, the observations before and after the changepoint follow different probability distributions \citep{chen2012parametric, killick2014changepoint}. We can use this information for prediction, monitoring, and decision-making purposes. Changepoint estimation or changepoint mining is common in many fields such as financial time series data \citep{thies2018bayesian}, environmental studies \citep{reeves2007review}, genome research \citep{muggeo2011efficient}, signal processing \citep{lavielle2005using}, quality control \citep{lai1995sequential}, and medical research \citep{bosc2003automatic}. During the 1950s, \cite{page1955test} first proposed a methodology for detecting only one change in a one-parameter (location-type) model. For a univariate Gaussian distribution, \cite{chernoff1964estimating} and \cite{gardner1969detecting} studied changepoint detection for the mean component, and \cite{hsu1977tests} studied a similar problem for the variance. While their studies were limited to a single parameter and a single changepoint scenario, \cite{hawkins1992detecting} proposed a novel approach for detecting a single shift in any known function of the unknown mean and covariance of an arbitrary multivariate distribution, and \cite{inclan1993detection} proposed a multiple changepoint detection procedure for the variance of a Gaussian distribution using posterior odds in a Bayesian setting. \cite{lerman1989detecting} discussed techniques and results on changepoint estimation for the correlation coefficient of the bivariate Gaussian distribution. 

 There are various testing procedures available in the literature for the purpose of changepoint estimation. One of the most common and earliest proposed choices is the CUmulative SUM (CUSUM) procedure \citep{page1954continuous}. CUSUM is particularly useful for detecting changes in a process over time, such as changes in the mean, variance, or distribution of the data. It is widely used in quality control, signal processing, and environmental monitoring, among other fields, to identify when a process has changed significantly from its previous state. Besides, certain moving window based methods like MOSUM \cite{eiauer1978use} have also been proposed. \cite{gurevich2005change} proposed generalized maximum likelihood asymptotic power one tests that aim to detect a single changepoint in the parameters of a logistic regression model. \cite{gurevich2010retrospective} considered distribution free generalized changepoint detection policies where the data distribution under the null hypothesis is not necessarily the same with the data distribution before the changepoint under the alternative. \cite{vexler2006guaranteed} proposed and examined a class of generalized maximum likelihood asymptotic power one tests for detection of various types of changes in a linear regression model. The proposed retrospective tests are based on martingales structured Shiryayev–Roberts statistics. \cite{vexler2009optimal} showed that retrospective change point detection policies based on Shiryayev–Roberts statistics are non-asymptotically optimal in the context of most powerful testing. A detailed discussion on parametric and non-parametric changepoint detection approaches is in \cite{csorgo1997limit} and \cite{chen2012parametric}.

While the literature on changepoint estimation in the context of EVT is scarce, the most common choice of the testing procedure is the Likelihood Ratio Test (LRT). In the context of changepoint estimation for a sequence of extremes, \cite{leadbetter1983extremes} and \cite{embrechts1997modelling} studied some theoretical properties of some test statistics. In a book chapter, \cite{da2004change} briefly discussed LRT for analyzing changes in the extreme value copulas and illustrated it for the bivariate Gumbel copula. \cite{jaruvskova2008analysis} discussed LRT for detecting a changepoint in the location parameter of annual maxima and minima series and described some methods for finding critical values. Instead of focusing on block maxima, \cite{dierckx2010changepoint} discussed LRT for detecting changes in the shape parameter of the generalized Pareto distribution, which is the only possible limiting distribution of high-threshold exceedances. While the previous approaches focused on a frequentist analysis of extreme values, \cite{nascimento2017bayesian} discussed a Bayesian method to identify changepoints in a sequence of extremes.  \cite{de2020tracking} proposed a time-varying extreme value copula where the authors model its angular surface using Bernstein polynomials and allow for changepoints in the parameters of the copula model. For detecting changepoints in financial extremes, \cite{lattanzi2021change} discussed a Bayesian approach for analyzing threshold exceedances using a generalized Pareto distribution while modeling the bulk of the distribution using a finite mixture of gamma distributions.

Apart from the context of EVT, \cite{worsley1979likelihood} examined the utilization of LRT to identify shifts or changes in the location of normal populations, while \cite{worsley1983power} focused on studying the power of LRT for detecting changes in binomial probabilities. Later, \cite{srivastava1986likelihood} investigated LRT for identifying changepoints in a multivariate setting, and \cite{asoka2003epidemic} discussed the detection of epidemic changes using LRT for exponential distribution. Further, \cite{zhao2013changepoint} studied detecting changepoints in two-phase linear regression models using LRT, and \cite{said2017likelihood} used LRT for detecting changepoints in a sequence of univariate skew-normal distributed random variables. Without imposing any parametric model assumption, \cite{zou2007empirical} proposed a changepoint detection procedure based on the empirical likelihood, and its asymptotic properties are justified by the work of \cite{owen1988empirical}. As an alternative to LRT, \cite{chen1997testing}, \cite{chen2006information}, \cite{hasan2014information}, \cite{ngunkeng2014information}, and \cite{cai2016changepoint} used the information criteria approach for changepoint detection. Specifically, \cite{chen2006information} proposed Modified Information Criterion (MIC) and showed that MIC has simple asymptotic behaviors and is consistent. While LRT generally performs better in detecting changepoints occurring near the middle of the data sequence, MIC includes some correction terms that allow better performance in detecting changepoints occurring at the very beginning or at the very end of the data. Among its applications, \cite{said2019modified} and \cite{tian2023changepoint} used MIC for changepoint estimation in the case of skew-normal and Kumaraswamy distributions, respectively.

Similar to the Pearson's correlation coefficient for analyzing dependence in the bulk of the joint distribution of two random variables, the strength of the dependence in tails (both lower and upper) is measured using extremal dependence or $\chi$-measure. Our purpose is to identify structural changes in the $\chi$-measure between the daily stock rate of return (RoR) of the two largest Indian airlines, IndiGo and SpiceJet, during the COVID-19 pandemic in 2019. To accomplish this, we discuss necessary preprocessing steps, including transforming the daily maximum and minimum RoR series so that each of them for IndiGo and SpiceJet jointly follow the BHR distribution. The main objective of this paper is to estimate the point at which a change occurs in the $\chi$-measure of the BHR distribution  separately for the upper and lower tails. Given the one-to-one correspondence between the dependence-related parameter of the BHR distribution and its $\chi$-measure, we explore a changepoint detection procedure utilizing LRT and MIC to achieve our goal. To assess the performance of these methods in practical settings with limited data samples, we numerically investigate their effectiveness. Since closed-form expressions for the finite sample distributions of LRT and MIC do not exist, we derive critical values and assess the power of the hypothesis testing problem across a range of low to high values of $\chi$-measure. In our data application, we explore the likelihood of different changepoints in the upper and lower tails from the beginning of the COVID-19 pandemic until the end of the third wave based on both LRT and MIC.

We organize the paper as follows. In Section \ref{sec:background}, we discuss a summary of univariate and bivariate EVT, quantification of extremal dependence, and the definition of the BHR distribution. Section \ref{sec:Exploration_and_Data_Preprocessing} discusses the Indian aviation stock price dataset and some exploratory analyses. We describe LRT and MIC for detecting the most crucial changepoint in a sequence of BHR-distributed bivariate random vectors in Section \ref{sec:Methodology}. Section \ref{sec:Simulation} discusses results on critical values and power comparison between LRT and MIC based on an extensive simulation study. In Section \ref{sec:Data_Application}, we apply our methodology for joint analysis of the stock prices of SpiceJet and IndiGo during the COVID-19 pandemic. Section \ref{sec:Conclusion} concludes and discusses some scopes for future research.


\section{Background on Extreme Value Theory}
\label{sec:background}

In this section, we summarize the theoretical background of univariate and bivariate Extreme Value Theory (EVT); they justify the methodology we adopt for data pre-processing as well as drawing inferences. Subsequently, before we discuss the changepoint estimation for the bivariate H\"usler-Reiss (BHR) distribution, we discuss how we obtain this distribution as a special case of a bivariate generalized extreme value distribution in Section \ref{subsec:bivariate_hr}; this justifies the importance of using the BHR distribution for drawing tail inferences in our context. We also briefly summarize the key measures of dependence in EVT that are necessary for understanding the features in the tail of the data. In the next section, we fit the univariate generalized extreme value distributions discussed here and explore the dependence in the tails of the data empirically using the measures described here.

 A brief summary and motivation of the technical details in Sections \ref{subsec:univariate_bivariate} and \ref{subsec:bivariate_hr} are as follows. The central limit theorem says that for any distributional assumption of the underlying sequence of random variables (univariate), as long as the underlying regularity conditions hold, the only possible asymptotic distribution of the sample mean is normal. Similarly, the Fisher-Tippett theorem says that for any distributional assumption of the underlying sequence of random variables, as long as the underlying regularity conditions hold and some non-degenerate limiting distribution exists, the only possible distribution of the block maxima/minima is the generalized extreme value distribution. In practice, we often observe only the sample mean (like average daily air temperature, instead of a full curve of air temperature throughout 24 hours) and hence a standard assumption is normality. Similarly, here we observe the daily maximum/minimum rate of return instead of the full curve of stock prices. At least we do not have access to such functional data. Hence, a natural choice of the distribution of daily maximum/minimum rate of return is the generalized extreme value distribution. When we have bivariate observations (obtained from bivariate curves with components corresponding to two airlines), the natural choice is the bivariate generalized extreme value distribution. Unlike a bivariate normal distribution, a bivariate generalized extreme value distribution can have different types of dependence structures. In practice, it is often difficult, if not impossible, to judge which class we should pick and the common choice is picking a specific class and draw inference based on that. Besides, there is a well-known result that if we assume the bivariate curves to be Gaussian processes, which is often a standard assumption for continuous stochastic processes across scientific disciplines, the only possible case of bivariate generalized extreme value distribution is the BHR distribution. Hence, we model the daily maximum/minimum rate of return for the two airlines using that distribution.

In multivariate extreme value theory, the dependence between two variables are often measured in terms of $\chi$-measure, which is essentially the (limiting) conditional probability of one variable being extremely large (small) given the other variable is also extreme large (small). The difference with Pearson's correlation is that $\chi$-measure does not measure linear association like Pearson's correlation and purely focuses on the tails. Hence, the bulk of the data do not interfere with drawing inferences about the strength of the dependence between two variables in the tails. Because of our focus on assessing the dependence structure within the extremes, i.e., the components of daily maximum/minimum rate of return series, drawing inferences based on $\chi$-measure is more appropriate than that based on Pearson's correlation.

\subsection{Univariate and Bivariate Extreme Value Distributions}
\label{subsec:univariate_bivariate}
The EVT deals with \textit{extreme} observations; they are usually defined as block maxima/minima (e.g., annual maxima/minima of daily observations) or threshold exceedances (e.g., the observations above the $0.98^{th}$ data quantile or below the $0.02^{th}$ data quantile). We stick to the first type of definition in our case. Under both definitions, the important aspect of EVT lies in describing the tail behavior of a stochastic process. In mathematical notations, consider $X_1, X_2, \ldots$, a sequence of independent random variables, where $X_i$'s are measured on a regular time scale, following a common continuous cumulative distribution function (CDF) $F_0(\cdot)$. The main goal of EVT is to study the asymptotic behavior of $M_n=\max_n \{X_1, X_2, \ldots, X_n\}$. The CDF of $M_n$ is given by $\textrm{P}(M_n \leq z)=F_0(z)^n$. But, in practice, $F_0(\cdot)$ is generally unknown. One approach would be estimating it from observed data in case the full sequence of observations is available; however, a small overestimation or underestimation usually influences the final inference substantially. Here, the asymptotic behavior of $\tilde{M}_n=\min_n \{X_1, X_2, \ldots, X_n\}$ is similar to that of $M_n$ as we can write $\tilde{M}_n=-\max_n \{-X_1, -X_2, \ldots, -X_n\}$. Thus, we focus on discussing the asymptotic behavior of $M_n$ only.

In this context, the celebrated Fisher-Tippett theorem \citep{fisher1928limiting} states that for any arbitrary CDF $F_0(\cdot)$, if there exist sequences of real numbers $\{a^0_n\}$ and $\{b^0_n\}$ and a non-degenerate CDF $G_0(\cdot)$ satisfying $\lim_{n\rightarrow\infty} F_0^n(a^0_nx+b^0_n)=G_0(x)$ pointwise, then $G_0(\cdot)$ must belong to either Gumbel, Fr\'echet, or Weibull families. Here, $F_0(\cdot)$ is said to belong to the \textit{Domain of Attraction} of $G_0(\cdot)$, and $a^0_n\in\mathbb{R}^+$ and $b^0_n\in\mathbb{R}$ are normalizing constants. Combining these three categories, $G_0(\cdot)$ is the CDF of the generalized extreme value (GEV) distribution given by
\begin{equation}\label{eq:gev}
G_0(y)\;=\;\exp\left[-\left\{1\;+\;\xi\left(\frac{y-\mu}{\sigma}\right)\right\}_+^{-1/\xi}\right],   
\end{equation}
where $\mu \in \mathbb{R}$, $\sigma \in \mathbb{R}^+$, and $\xi \in \mathbb{R}$ are the location, scale, and shape parameters of the GEV distribution, and $x_+=\max\{x,0\}$. Depending on whether the shape parameter $\xi$ is zero, positive, or negative, the GEV family is called the Gumbel family, Fr\'echet family, or Weibull family, respectively. We denote the distribution with CDF $G_0(\cdot)$ by $\textrm{GEV}(\mu, \sigma, \xi)$, and in practice, we assume $M_n \sim \textrm{GEV}(\mu, \sigma, \xi)$ in a limiting sense. The centering (through $b^0_n$) and scaling (through $a^0_n$) of $M_n$ in the Fisher-Tippett theorem are assumed to be adjusted by the parameters $\mu$ and $\sigma$, respectively.

Multivariate extreme value theory focuses on the case where multiple sequences of random variables are available and we are interested in assessing the joint asymptotic behavior of the block maxima for all sequences. For our purpose, we stick to the bivariate extremes scenario. Suppose $\{(X_1, Y_1),(X_2, Y_2),\ldots \}$ is a sequence of IID bivariate random vectors having a common continuous CDF $F(\cdot, \cdot)$. The classical theory for characterizing the extremal behavior of bivariate extremes is based on the asymptotic behavior of the component-wise block maxima vector $\bm{M}_n=(M_{X,n}, M_{Y,n})'$, where $M_{X,n}=\max_{1\leq i\leq n}\{X_i\}$ and $M_{Y,n}=\max_{1\leq i\leq n}\{Y_i\}$. Extending the Fisher-Tippett theorem for bivariate cases, \cite{campbell1973asymptotic} stated that if there exist real sequences $\{a_n\}$, $\{b_n\}$, $\{c_n\}$ and $\{d_n\}$, where $a_n, c_n \in \mathbb{R}^+$ for all $n$, and a bivariate non-degenerate CDF $G(\cdot,\cdot)$ that satisfy $\lim_{n\rightarrow\infty}F^n(a_nx+b_n,c_ny+d_n)=G(x,y)$ pointwise, then $G(\cdot,\cdot)$ is called the bivariate GEV distribution. Here, the standard Fisher-Tippett theorem applies to both $\{X_1, X_2,\ldots \}$ and $\{Y_1, Y_2,\ldots \}$, and the limiting CDFs of renormalized $M_{X,n}$ and $M_{Y,n}$ are of the form \eqref{eq:gev}. Suppose, $M_{X,n} \sim \textrm{GEV}(\mu_X, \sigma_X, \xi_X)$ and $M_{Y,n} \sim \textrm{GEV}(\mu_Y, \sigma_Y, \xi_Y)$ in a limiting sense. The CDF $G(\cdot,\cdot)$ can be written as
\begin{equation} \label{eq:cdf_to_exponent}
G(x,y)=\exp\{-V(\widetilde{x},\widetilde{y})\}, 
\end{equation}
where $\widetilde{x}=[1+\xi_X (x-\mu_X) / \sigma_X]^{1/\xi_X}$ and $\widetilde{y}=[1+\xi_Y (y-\mu_Y) / \sigma_Y]^{1/\xi_Y}$. Here, we assume $[1+\xi_X(x-\mu_X)/\sigma_X]>0$, $1+\xi_Y(y-\mu_Y)/\sigma_Y>0$ and $V(\widetilde{x},\widetilde{y})=2\int_0^1\max\left\lbrace w/\widetilde{x}, (1-w)/\widetilde{y}\right\rbrace \mathrm{d}\widetilde{G}(w)$, where $\widetilde{G}(\cdot)$ is a CDF on $[0,1]$ satisfying the mean constraint $\int_0^1 w \mathrm{d}\widetilde{G}(w)=1/2$. Here, $V(\cdot,\cdot)$ is called the exponent measure of $G(\cdot, \cdot)$. 

\subsection{Extremal dependence and F-madogram}
\label{subsec:ext_dependence}
For certain bivariate CDFs with finite second moments, even if the two components of the corresponding bivariate random vector are highly correlated, i.e., the dependence in the bulk of the joint distribution is strong, the dependence in the tails can be weak or negligible. Similar to Pearson’s correlation for analyzing dependence in the bulk of the joint distribution, the most common metric for measuring dependence in the tails is called extremal dependence or $\chi$-measure introduced by \cite{sibuya1960bivariate}. This measure does not require the second moments to be necessarily finite.

For a bivariate random vector with components $X$ and $Y$, and marginal CDFs $F_X(\cdot)$ and $F_Y(\cdot)$, the upper and lower tail $\chi$-measures at a quantile level $u$ are defined as
\begin{eqnarray} \label{eq:chi_u}
    \nonumber \chi_U(u) &=& \textrm{P}\{F_Y(Y)>u|F_X(X)>u\}, \\
    \chi_L(u) &=& \textrm{P}\{F_Y(Y)<u|F_X(X)<u\},~~~~u \in (0,1),
\end{eqnarray}
while the limiting $\chi$-measures are defined as $\chi_U = \lim_{u \uparrow 1} \chi_U(u)$ and $\chi_L = \lim_{u \downarrow 0} \chi_L(u)$. Here, $\chi_U(u)$ and $\chi_L(u)$ are not uniquely defined and dependent on $u$. Thus, for a unique measure of extremal dependence, we use $\chi_U$ and $\chi_L$ unless specified. Intuitively, a high value of $\chi_U$ indicates the tendency of $Y$ being extremely large given $X$ is extremely large, and similarly, a high value of $\chi_L$ indicates the tendency of $Y$ being extremely small given $X$ is extremely small. If $\chi_U \in (0,1]$, we call $X$ and $Y$ to be asymptotically dependent in the upper tail, while for $\chi_U = 0$, $X$ and $Y$ are said to be asymptotically independent in the upper tail. Similarly, if $\chi_L \in (0,1]$, we call $X$ and $Y$ to be asymptotically dependent in the lower tail, while for $\chi_L = 0$, $X$ and $Y$ are said to be asymptotically independent in the lower tail. In \eqref{eq:chi_u}, $X$ and $Y$ are interchangeable. More details are in \cite{coles2001introduction}.

For a bivariate GEV distribution, \eqref{eq:cdf_to_exponent} and \eqref{eq:chi_u} are linked through the equation 
\begin{equation}\label{eq:chiu_to_exponent}
\chi_U(u)\;=\;\chi_L(u)\;=\;2\;-\;V(1,\;1),
\end{equation}
uniformly for $u \in (0,1)$, and thus, $\chi=2-V(1,1)$; see Chapter 8 of \cite{coles2001introduction}. In case IID replications of $(X, Y)'$ are available, $\chi(u)$ can be computed empirically. However, the concept of $\chi$-measure can be extended from a bivariate setting to a stochastic process setting. For a stationary extremal time series $\{ Z_1, Z_2, \ldots \}$ with a common marginal CDF $F(\cdot)$, the $\chi$-measure at a temporal lag $h=1,2,\ldots$ can be investigated using $F$-madogram \citep{cooley2006variograms} given by $\nu_h=E\left[\vert F(Z_{t+h})-F(Z_t) \vert \right]/2$ for any $t$. Then, the $\chi$-measure at lag $h$, say $\chi_h$, satisfies
\begin{equation}
\chi_h = 2 - \left(\frac{1+2\nu_h}{1-2\nu_h}\right).
\label{eq:chi_to_madogram}
\end{equation}
In practice, for testing temporal extremal independence, we calculate $\chi_h$ empirically and check whether the values are close to zero. If $\chi_h$ are negligible for all $h$, we can safely ignore extremal dependence and model $\{ Z_1, Z_2, \ldots \}$ as IID observations.

\subsection{Bivariate H\"usler-Reiss Distribution}
\label{subsec:bivariate_hr}
For a sequence of independent random variables $X_1, X_2, \ldots$ measured on a regular time scale and with a common CDF $F_0(\cdot)$, if $F_0(\cdot)=\Phi(\cdot)$--the standard Gaussian CDF, $G_0(\cdot)$ in \eqref{eq:gev} belongs to the Gumbel family \citep{david2004order}, i.e., $\xi = 0$, which is defined in a limiting sense $\xi \rightarrow 0$, and we have $G_0(y)=\exp(-\exp[-y]),~y\in \mathbb{R}$. Here, for a sequence $\{b_n\}$ satisfying $b_n = n\phi(b_n)$, where $\phi(\cdot)$ denotes a standard Gaussian density, we have $\lim_{n \uparrow \infty} \Phi^n\left(b_n+x/b_n\right) = \exp(-\exp[-x])$, the standard Gumbel distribution, for all $x \in \mathbb{R}$. Similarly, for a sequence of bivariate random vectors $\{(X_1, Y_1),(X_2, Y_2),\ldots \}$ with common CDF $F$, if $F = \Phi_{\rho}$--the bivariate standard Gaussian CDF with correlation $\rho$, the component-wise renormalized block maxima $M_{X,n}$ and $M_{Y,n}$ (with notations as in Section \ref{subsec:univariate_bivariate}) follow the standard Gumbel distribution. Similarly, the component-wise renormalized block minima $\tilde{M}_{X,n} = \min_{1\leq i\leq n}\{X_i\}$ and $\tilde{M}_{Y,n}= \min_{1\leq i\leq n}\{Y_i\}$ also follow the same distribution.

For the bivariate Gaussian distributions, \cite{sibuya1960bivariate} proved that the components of the corresponding random vector are asymptotically independent, i.e., $\chi = 0$, for any value of the correlation coefficient $\rho$ less than one. In this context, \cite{Husler1989maxima} suggested an asymptotic formulation where the correlation coefficient $\rho\equiv\rho_n$ of a bivariate Gaussian distribution varies as sample size $n$ increases, and they proved that the marginal maxima/minima are neither asymptotically independent nor completely independent if $(1-\rho_n)\log(n)$ converges to a positive constant as $n\uparrow\infty$; the limiting joint distribution is called the bivariate H\"usler-Reiss (BHR) Distribution. More specifically, \cite{Husler1989maxima} proved that if $\lim_{n \uparrow \infty} (1-\rho_n)\log(n) = \Lambda^{-2}\in[0,\infty]$, then $\forall~x,y \in \mathbb{R}$, $\lim_{n \uparrow \infty} \Phi_{\rho_n}^n\left(b_n+x/b_n,b_n+y/b_n\right) = H_{\Lambda}(x,y)$, where 
\begin{equation}
    H_{\Lambda}(x,y)=\exp\left[-\exp(-x) \,\Phi\left\lbrace \frac{1}{\Lambda}+\frac{\Lambda}{2} (y - x)\right\rbrace - \exp(-y)\,\Phi\left\lbrace\frac{1}{\Lambda}+\frac{\Lambda}{2} (x - y)\right\rbrace\right]
\label{eq:bhr_cdf}
\end{equation}
is the CDF of the BHR distribution with the dependence-related parameter $\Lambda$.

Obtaining the exponent measure $V(\cdot,\cdot)$ for the BHR distribution from \eqref{eq:cdf_to_exponent} and \eqref{eq:bhr_cdf} is straightforward. Further, from \eqref{eq:chiu_to_exponent}, we obtain the (limiting) extremal dependence measure to be $\chi=2-V(1,1)=2\overline{\Phi}(1/\Lambda)$, where $\overline{\Phi}(\cdot)$ denotes the standard Gaussian survival function. Here, $\chi$ is monotonically increasing with $\Lambda$, where $\Lambda=0$ and $\Lambda=\infty$ imply independence and complete dependence between the components. 

For likelihood-based testing procedures discussed in Section \ref{sec:Methodology}, the probability density function (PDF) is required. From \eqref{eq:bhr_cdf}, the corresponding PDF is given by
\begin{eqnarray}\label{eq:bhr_pdf}
\nonumber h_{\Lambda}(x,y)&=&\exp\left[-\left(\Psi(\Lambda)+\widetilde{\Psi}(\Lambda)\right)\right]\left[\frac{\Lambda}{2}\left\{\left(\psi(\Lambda)+\widetilde{\psi}(\Lambda)\right)+\frac{\Lambda}{2}
\left(\psi^{(1)}(\Lambda)+\widetilde{\psi}^{(1)}(\Lambda)\right)\right\}\right.\\
&& \left.+\left\{\Psi(\Lambda)+\frac{\Lambda}{2}\left(\psi(\Lambda)-\widetilde{\psi}(\Lambda)\right)\right\}
\left\{\widetilde{\Psi}(\Lambda)+\frac{\Lambda}{2}\left(\widetilde{\psi}(\Lambda)-\psi(\Lambda)\right)\right\}\right],
\end{eqnarray}
where
\begin{eqnarray}
\nonumber && \Psi(\Lambda)=e^{-x}\Phi\left(1/\Lambda+\Lambda(y-x)/2\right), ~~\widetilde{\Psi}(\Lambda)=e^{-y}\Phi\left(1/\Lambda+\Lambda(x-y)/2\right), \\
\nonumber && \psi(\Lambda)=e^{-x}\phi\left(1/\Lambda+\Lambda(y-x)/2\right),~~ \widetilde{\psi}(\Lambda)=e^{-y}\phi\left(1/\Lambda+\Lambda(x-y)/2\right), \\
\nonumber && \psi^{(1)}(\Lambda)=e^{-x}\frac{\partial}{\partial y}\left\{\phi\left(1/\Lambda+\Lambda(y-x)/2\right)\right\}, ~~\widetilde{\psi}^{(1)}(\Lambda)=e^{-y}\frac{\partial}{\partial y}\left\{\phi\left(1/\Lambda+\Lambda(x-y)/2\right)\right\}.
\end{eqnarray}

\begin{figure}[t!]
\includegraphics[height = 0.3\linewidth]{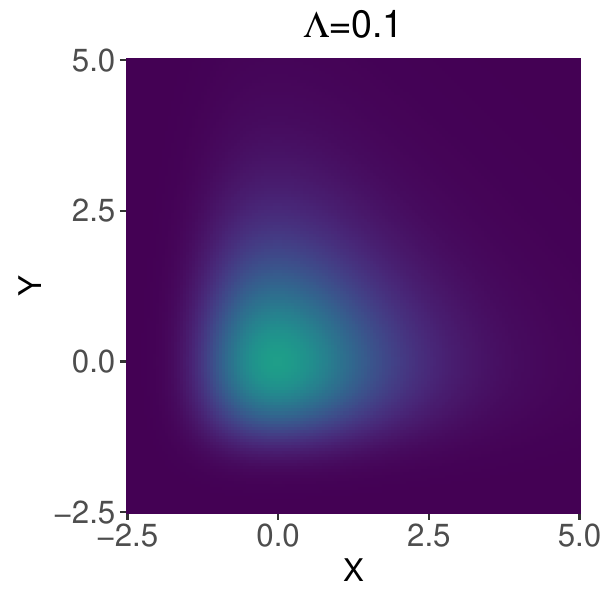}
\includegraphics[height = 0.3\linewidth]{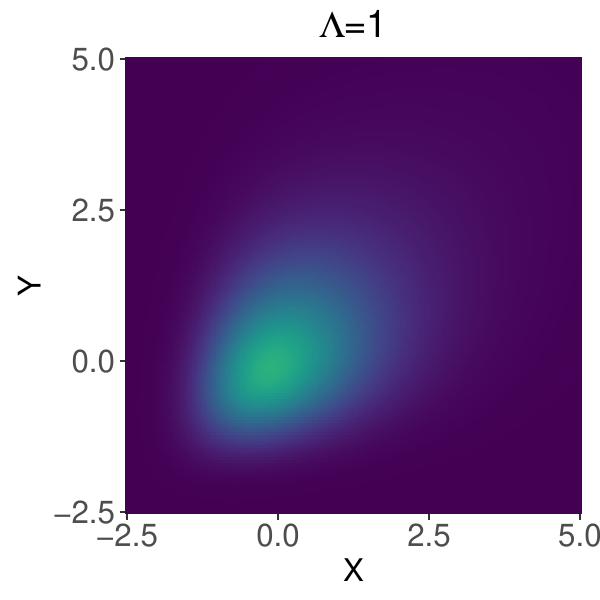}
\includegraphics[height = 0.3\linewidth]{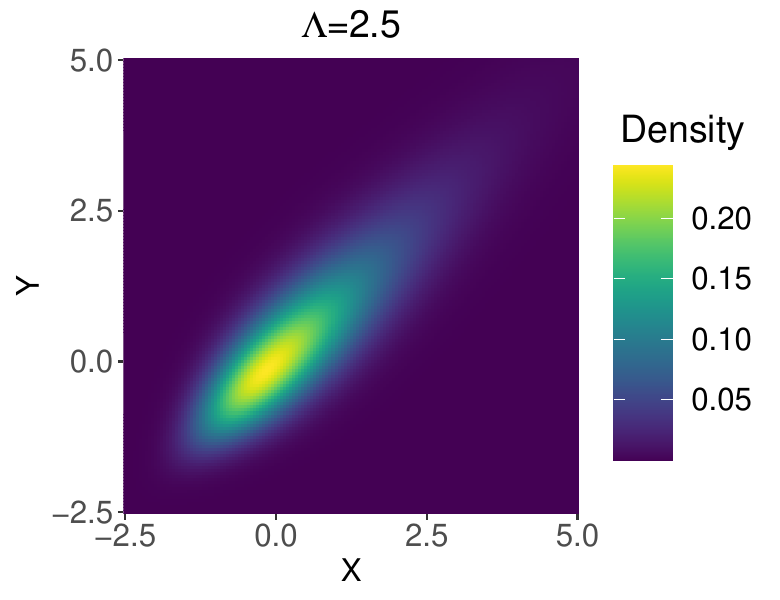}
\caption{Densities of the bivariate H\"usler-Reiss distribution for different values of the dependence-related parameter $\Lambda$. A large (small) value of $\Lambda$ induces a strong (weak) dependence.} 
\label{fig:bhr_pdf}
\end{figure}

For $\Lambda = 0.1, 1$, and $2.5$, we present the PDFs of the BHR distribution in Figure \ref{fig:bhr_pdf}. For these choices of $\Lambda$, the values of the $\chi$-measure are $1.52 \times 10^{-23}$, 0.3173, and 0.6892, respectively. The figure demonstrates that a large (small) value of $\Lambda$ induces a strong (weak) dependence. Interchanging the arguments ($x$ and $y$), the PDF \eqref{eq:bhr_pdf} remains the same; thus, the components of the BHR distribution are exchangeable.

\section{Aviation Stock Price Dataset: Exploration and Pre-Processing}
\label{sec:Exploration_and_Data_Preprocessing}
\subsection{Data Description}
This article focuses on analyzing the daily stock prices of two prominent Indian aviation companies, IndiGo and SpiceJet. We obtain the related datasets from the website \url{https://www.investing.com/}. We consider the period from December 2, 2019, to May 31, 2022, which coincides with the global COVID-19 pandemic. According to the World Health Organization (\url{https://covid19.who.int/region/searo/country/in}), India experienced three severe waves of the pandemic during this time. The first wave lasted from mid-March 2020 to mid-January 2021, followed by the deadliest second wave from mid-March 2021 to mid-August 2021. Lastly, the third wave persisted from mid-December 2021 to mid-March 2022. These waves had significant repercussions on various sectors of society and the economy, including the aviation market. Due to widespread travel restrictions and reduced investor activity, share prices experienced declines during the lockdown periods. At the same time, during unlock phases, share prices exhibited sudden increments. Our analysis aims to explore how the COVID-19 waves influenced the dependence between the share prices of these two aviation companies, with a focus on detecting the most crucial changepoint in each of their upper and lower tail dependences, due to their link with simultaneous sell and buy positions, respectively.

\subsection{Daily Maximum and Minimum Rates of Return}
Instead of dealing with the stock prices in their original scales, we focus on analyzing their rate of return (RoR), which are more meaningful comparison metrics. This rate signifies the percentage variation in the value of a particular stock investment over a specified period, indicating the level of gain or loss experienced by the investor on their stock holding. The RoR can be computed using the following formula:
\begin{equation}
\nonumber \textrm{RoR}\,=\,\frac{(\textrm{Ending price}\,-\,\textrm{Beginning price})
}{\textrm{Beginning price}}.
\label{returnrate}
\end{equation}
Here, the ending price corresponds to the stock price after a specified period, while the beginning price represents the stock price at the beginning of the same period. 
The RoR can take both positive and negative values, indicating whether an investor gains profit or incurs a loss on the investment, respectively. This measure holds great importance in evaluating the performance of a stock and comparing it to other investment options available. A standard period for calculating RoR is 24 hours and we stick to it here.

Given our focus on the right and left tails of the joint distribution of RoR, we define the daily maximum and minimum RoR as follows. Suppose $Y_{t}(s)$ represents the value of the stock on a specific day $t$ at time $s$. We calculate the daily maximum and minimum RoR on day $t$, respectively, by
\begin{eqnarray} \label{eq:return_rate}
\nonumber  R^{\textrm{max}}_t\;&=\;\max_{s,s^{'}}\;\left\{\dfrac{Y_{t+1}\,(s)\;-\;Y_{t}\,(s^{'})}{Y_{t}\,(s^{'})}\right\}=\;\dfrac{\max_{s}\,Y_{t+1}(s)}{\min_{s^{'}}Y_t\,(s^{'})}\,-\,1, \\
    R^{\textrm{min}}_t\;&=\;\min_{s,s^{'}}\;\left\{\dfrac{Y_{t+1}\,(s)\;-\;Y_{t}\,(s^{'})}{Y_{t}\,(s^{'})}\right\}=\;\dfrac{\min_{s}\,Y_{t+1}(s)}{\max_{s^{'}}Y_t\,(s^{'})}\,-\,1.
\end{eqnarray}

The metrics in \eqref{eq:return_rate} are not common in the financial or statistical literature. However, we provide a philosophical justification for the two metrics in the following. Sudden changes in the stock prices are often governed by news. For example, with the immediate announcement of lockdown during COVID-19 at a particular time of the day, numerous investors sold stocks immediately. The stock prices may recover slightly afterwards but the main impact on the market happens at the moment of broadcasting the news. Thus, choosing stock price as a function of time throughout the day and looking at the rate of change within a timeframe beginning on the $t$-th day and ending on the $(t+1)$-th day, we can have a better idea of the maximum or minimum possible variability within two consecutive days. A limitation of these metrics is that they do not explain maximum/minimum possible change within a single day.

\begin{figure}
\centering
\includegraphics[width=0.95\linewidth]{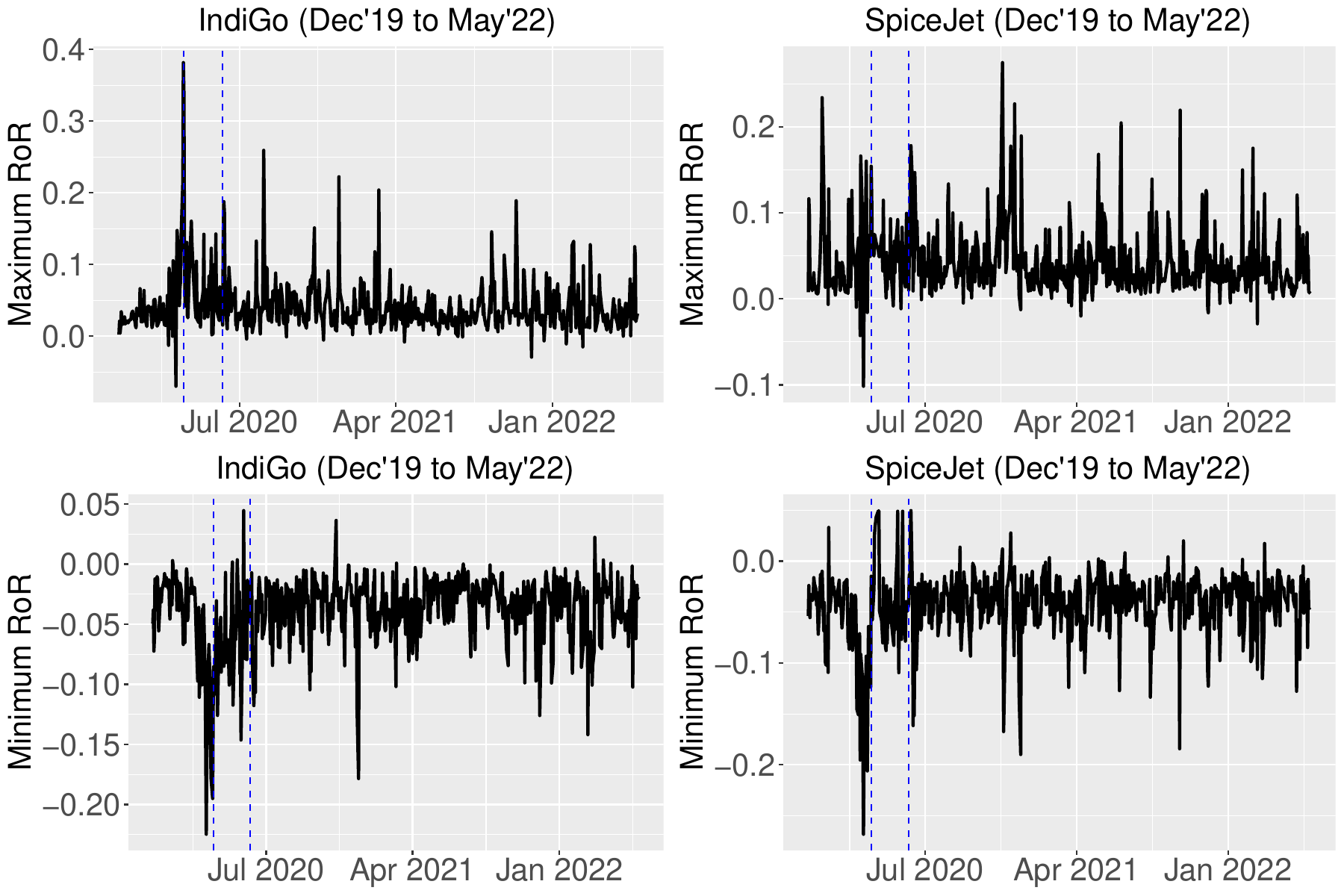}
\caption{Daily maximum (top) and minimum (bottom) rate of return series of IndiGo and SpiceJet airlines, as defined in \eqref{eq:return_rate}. The two vertical lines denote the two dates March 25, 2020, the day the first lockdown was declared and June 1, 2020, the day the first unlock phase was declared by the Government of India.}
\label{fig:returnrate_ts_maxmin}
\end{figure}

The standard publicly available datasets include both the daily maximum and minimum stock prices $\max_{s}\,Y_{t}(s)$ and $\min_{s}\,Y_{t}(s)$ for different $t$'s. Hence, despite having access to the curves $Y_{t}(\cdot)$ and $Y_{t+1}(\cdot)$, calculating $R^{\textrm{max}}_t$ and $R^{\textrm{min}}_t$ from them the available data is straightforward. The plots of $R^{\textrm{max}}_t$'s and $R^{\textrm{min}}_t$'s for both IndiGo and SpiceJet are presented in Figure \ref{fig:returnrate_ts_maxmin}. During the initial days of the COVID-19 pandemic in India, $R^{\textrm{max}}_t$ profiles attain negative values for both airlines; however, the lowest value of $R^{\textrm{max}}_t$ is more negative for SpiceJet than that for IndiGo, and the subsequent $R^{\textrm{max}}_t$ values attain higher positive values for IndiGo than that for SpiceJet. A clear nonstationary pattern is observable in both $R^{\textrm{max}}_t$ series. The $R^{\textrm{min}}_t$ profiles generally remain negative for IndiGo during the initial lockdown period, but it becomes positive for one instance, with a significant margin, for SpiceJet. Subsequently, for more than one month, $R^{\textrm{min}}_t$ attains negative values, with   attaining more negative values for SpiceJet compared to IndiGo. After that period, we observe fewer highly positive spikes in the $R^{\textrm{min}}_t$ profile for IndiGo than that for SpiceJet. The largest spike in maximum RoR for IndiGo matches with the date when the first lockdown was declared. On the other hand, several spikes in the minimum RoR profile visible for SpiceJet before the day the first unlock phase was declared by the Government of India.

\subsection{Data Preprocessing}
\label{subsec:data_preprocessing}
While the Fisher-Tippett theorem described in Section \ref{sec:background} assumes the underlying sequence of random variables to be independently and identically distributed, the result also holds in the case of dependent sequences under some additional regularity conditions explained in \cite{leadbetter1988extremal}. Here, the original stock prices across different timestamps $s$ on a certain day $t$, denoted by $Y_{t}(s)$ in \eqref{eq:return_rate}, are likely to be dependent. Assuming the regularity conditions of \cite{leadbetter1988extremal} hold, we assume that $R^{\textrm{max}}_t$'s follow the GEV distribution in \eqref{eq:gev}, characterized by three parameters-- location ($\mu$), scale ($\psi$), and shape ($\xi$). Because of not having access to the curves $Y_{t}(\cdot)$ and $Y_{t+1}(\cdot)$, only option is to assume that the regularity conditions of \cite{leadbetter1988extremal} hold and $R^{\textrm{max}}_t$'s and $R^{\textrm{min}}_t$'s follow the GEV distribution. However, such an assumption is common in the extreme value analysis literature. Therefore, to transform $R^{\textrm{max}}_t$'s to standard Gumbel margins, our first data preprocessing step involves estimating these parameters. However, the bottom panels of Figure \ref{fig:returnrate_ts_maxmin} show that assuming the GEV parameters to be constant across $t$ is questionable for both airlines. Hence, we assume the marginal model to be $R^{\textrm{max}}_t \sim \textrm{GEV}(\mu_t, \sigma_t, \xi_t)$ separately for each company. Here, we estimate the temporally-varying GEV parameters using local probability-weighted moments \citep[PWM,][]{hosking1985estimation} estimation. We repeat the same procedure for $R^{\textrm{min}}_t$'s.

In general, the term \textit{local likelihood}, introduced by \cite{tibshirani1987local}, refers to a method that estimates the parameters of a statistical model by considering local subsets of data instead of the entire dataset as a whole. The local likelihood approach involves placing a window around each observation and maximizing the likelihood function for the model's parameters within that window. In our case, we utilize a window size of 100 for each analysis and employ PWM estimation. Ideally, the window size can be considered to be a tuning parameter and chosen based on a thorough crossvalidation. While a too small value of the bandwidth might provide unreliable estimates, a too large value would oversmooth the parameter profiles. Unlike the traditional method of moments, where we equate population moments to their sample counterparts with equal weights, PWM utilizes a set of weights that depends on the probability density function of the population. These weights are carefully selected to capture the distribution's shape more accurately, and thus, the PWM method is used widely in extreme value analysis. Implementing local PWM estimation, we obtain the estimates of $\{\mu_t, \sigma_t, \xi_t\}$ separately for IndiGo and SpiceJet at each day $t$ during the COVID-19 pandemic. We obtained the shape parameter estimates for both $R^{\textrm{max}}_t$'s and $R^{\textrm{min}}_t$'s to be close to zero in almost all the cases, along with high standard errors. Thus, the shape parameter can be safely fixed to zero, i.e., the corresponding GEV distribution can be assumed to belong to the Gumbel class, for both airlines. We then apply the local PWM estimation procedure after setting $\xi_t = 0$ for all cases. 

Further, suppose we denote the joint daily maximum rate of return vector for day $t$ by $\bm{R}^{\textrm{max}}_t = (R^X_t, R^Y_t)'$, where $R^X_t$ and $R^Y_t$ denote the individual daily maximum rate of return on day $t$ for IndiGo and SpiceJet, respectively. Besides, let the corresponding local PWM estimates for the parameters of the underlying Gumbel distributions be given by $\{\widehat{\mu}^X_t, \widehat{\sigma}^X_t \}$ and $\{\widehat{\mu}^Y_t, \widehat{\sigma}^Y_t\}$, respectively. We transform $R^X_t$ and $R^Y_t$ to standard Gumbel margins following the location-scale transformations
\begin{equation}\label{eq:standard_gumbel}
    \widetilde{R}^X_t = \frac{R^X_t - \widehat{\mu}^X_t}{\widehat{\sigma}^X_t}, \quad \widetilde{R}^Y_t = \frac{R^Y_t - \widehat{\mu}^Y_t}{\widehat{\sigma}^Y_t},
\end{equation}
and obtain the transformed random vectors $\widetilde{\bm{R}}^{\textrm{max}}_t = (\widetilde{R}^X_t, \widetilde{R}^Y_t)$. We repeat the same procedure for $R^{\textrm{min}}_t$'s and denote the joint daily minimum rate of return vector transformed to standard Gumbel margins by $\widetilde{\bm{R}}^{\textrm{min}}_t$'s.

We explore the temporal extremal dependence within the time series $\{ \widetilde{R}^X_t \}$ and $\{ \widetilde{R}^Y_t \}$ for both $\widetilde{\bm{R}}^{\textrm{max}}_t$'s and $\widetilde{\bm{R}}^{\textrm{min}}_t$'s through $F$-madogram as in \eqref{eq:chi_to_madogram}. For all series, we observe that the estimated temporal $\chi$-measures are close to zero across low through high lags, and thus, we safely assume each of $\{ \widetilde{R}^X_t \}$ and $\{ \widetilde{R}^Y_t \}$ to be independent across $t$ for both $\widetilde{\bm{R}}^{\textrm{max}}_t$'s and $\widetilde{\bm{R}}^{\textrm{min}}_t$'s. To explore the extremal dependence between the components of $\widetilde{\bm{R}}^{\textrm{max}}_t$'s and $\widetilde{\bm{R}}^{\textrm{min}}_t$'s, we use $F$-madogram similar to its use for exploring temporal extremal dependence. We obtain the empirical $\chi$-measure for $\widetilde{\bm{R}}^{\textrm{max}}_t$'s and $\widetilde{\bm{R}}^{\textrm{min}}_t$'s to be 0.3516 and 0.3914, respectively. Under the null hypothesis of independence between the components of $\widetilde{\bm{R}}_t$, the bootstrap-based critical value is 0.0646 based on a test of level 0.95. Thus, we can safely claim that there is a strong extremal dependence between the components of each $\widetilde{\bm{R}}^{\textrm{max}}_t$'s and $\widetilde{\bm{R}}^{\textrm{min}}_t$'s, and further, we assume that the joint distribution of the components of each $\widetilde{\bm{R}}^{\textrm{max}}_t$'s and $\widetilde{\bm{R}}^{\textrm{min}}_t$'s to be the BHR distribution with CDFs of the form \eqref{eq:bhr_cdf}.

\section{Methodology}
\label{sec:Methodology}
In this section, we discuss the Likelihood Ratio Test (LRT) and Modified Information Criterion (MIC) to detect changepoints in the BHR distribution. From the data, we can estimate the model parameters based on different possible choices of the changepoint and subsequently, after plugging in the estimates within the likelihood function, we can calculate $Z'_T$ and $S'_T$. The tests are not distribution free and in a finite-sample setting, they depend on sample sizes. However, as long as we can obtain the critical values empirically, we can use these tests. In Section \ref{subsec:critical_vals}, we discuss obtaining the critical values empirically. This strategy in the context of using LRT and MIC for changepoint estimation is not new. For example, \cite{tian2022changepoint} and \cite{said2019modified} follow the same strategy for obtaining the critical values in the context of using LRT and related tests.

By an abuse of notation, we denote both $\widetilde{\bm{R}}^{\textrm{max}}_t$ and $\widetilde{\bm{R}}^{\textrm{min}}_t$ by a generic notation $\widetilde{\bm{R}}_t$ in this section. Let the transformed daily maximum/minimum rate of return vectors $\{ \widetilde{\bm{R}}_t = (\widetilde{R}^X_t, \widetilde{R}^Y_t), t=1, \ldots, T \}$ in \eqref{eq:standard_gumbel} be a sequence of independent observations from the BHR distribution with dependence-related parameters $\{\Lambda_t, t = 1, \ldots, T\}$ respectively. We are interested in testing for changes in the parameters $\Lambda_t$'s, i.e., our null and alternative hypotheses of interest are
\[H_{0}\;:\;\Lambda_{1}=\cdots =\Lambda_{T}=\Lambda, \quad H_{A}\;:\;\Lambda_{1}=\ldots = \Lambda_{\tau}\neq\Lambda_{\tau+1} = \cdots =\;\Lambda_{T-1}\;=\;\Lambda_{T} \]
for a changepoint $\tau$, if exists. Then, under $H_0$, the log-likelihood is
\begin{equation}
\label{log_likelihood_H0}
\begin{aligned}
&\log L_{H_0}(\Lambda)=\sum_{t=1}^T \log \left[ h_{\Lambda}\left(\widetilde{R}^X_t, \widetilde{R}^Y_t \right) \right],& 
\end{aligned}
\end{equation}
where $h_\Lambda(\cdot, \cdot)$ is as in \eqref{eq:bhr_pdf}, and the corresponding maximum likelihood estimate (MLE) of $\Lambda$, say $\widehat{\Lambda}$, is obtained by solving the following score equation 
\begin{equation}
    \frac{\partial}{\partial\Lambda}(\log\,L_{H_0}(\Lambda))=0.
    \label{score_eq_H0}
\end{equation}
Under the alternative hypothesis $H_{A}$, the log-likelihood function is
\begin{equation} 
\label{log_likelihood_HA}
\begin{aligned}
&\log L_{H_A}(\Lambda_1,\Lambda_T,\tau)= \sum_{t=1}^\tau \log \left[ h_{\Lambda_1}\left(\widetilde{R}^X_t, \widetilde{R}^Y_t \right)\right] + \log \left[ \sum_{t=\tau+1}^T h_{\Lambda_T}\left(\widetilde{R}^X_t, \widetilde{R}^Y_t \right) \right],
\end{aligned} 
\end{equation}
and we obtain the MLEs of $\Lambda_1$ and $\Lambda_T$, say $\widehat{\Lambda}_1$ and $\widehat{\Lambda}_T$ respectively, by solving the following score equations 
\begin{equation}
\label{score_eq_HA}    
\frac{\partial}{\partial\Lambda_1}(\log\,L_{H_A}(\Lambda_1,\Lambda_T,\tau))=0, \quad \frac{\partial}{\partial\Lambda_T}(\log\,L_{H_A}(\Lambda_1,\Lambda_n,\tau))=0.    
\end{equation}
The above score equations \eqref{score_eq_H0} and \eqref{score_eq_HA} require nonlinear optimization and we use the function \texttt{fbvevd} from the \texttt{R} package \texttt{evd} \citep{stephenson2002evd} to get the MLEs.

\subsection{Likelihood Ratio Test (LRT)}
The most commonly used test for changepoint detection problems is LRT. Here we assume that the underlying data are independent across time.
It works in the following way. Consider a timepoint $\tau$ between $1$ and $T$ at which a change occurs. We reject our null hypothesis, i.e., no changepoint, if we observe a high value of the (2-times the log) likelihood ratio, for a fixed $\tau$, given by 
\begin{equation} \label{eq:loglik_diff} 
\begin{aligned} 
\textrm{LR}(\tau)&=-2 \left\lbrace \log L_{H_0}(\widehat{\Lambda})-\log L_{H_A}(\widehat{\Lambda}_1,\widehat{\Lambda}_T)\right\rbrace,
\end{aligned}
\end{equation} 
where $\log L_{H_0}$ and $\log L_{H_A}$ are given by \eqref{log_likelihood_H0} and \eqref{log_likelihood_HA}, respectively. Further, considering a range of possible values of $\tau$, the alternative hypothesis is preferred if $\textrm{LR}(\tau)$ is high for any single $\tau$ and the related LRT statistic is given by
\begin{equation}
\nonumber Z_T\;=\;\max_{1 \leq \tau < T}\; \textrm{LR}(\tau). 
\end{equation}
If $\tau$ is small, we do have not sufficient data to obtain the MLE $\widehat{\Lambda}_1$. Similarly, for $\tau$ close to $T$, $\widehat{\Lambda}_T$ would be unstable due to insufficient data. Thus, to avoid high uncertainty of the estimates, \cite{ngunkeng2014information} suggested the modified version of $Z_T$, or in other words the trimmed LRT statistic $Z'_T$ given by
\begin{equation}
Z'_T\;=\;\max_{\tau_0<\tau<T-\tau_0}\; \textrm{LR}(\tau),\;\;\textrm{where}\;\;\tau_0=2 \lfloor \log(T) \rfloor. 
\label{modified_Zt}
\end{equation}
while other choices of $\tau_0$ have been proposed in the literature, e.g., \cite{liu2009changepoint} suggested $\tau_0\;=\;\lfloor \log\,T \rfloor^2$, we stick to the choice in \eqref{modified_Zt}. Once we reject our null hypothesis, the estimated changepoint is
\begin{equation}
\widehat{\tau}_{\textrm{LRT}}\;=\;\argmax_{\tau_0<\tau<T-\tau_0} \;\textrm{LR}(\tau).
\label{tauhat_LRT} 
\end{equation}
For a given level of significance $\alpha$, we reject our null hypothesis $H_0$ if $Z'_T>c_{\alpha,\;T}$, where $c_{\alpha,\;T}$ is the corresponding critical value. 
In Section \ref{sec:Simulation}, we numerically obtain $c_{\alpha,\;T}$ for different choices of $T$ and $\alpha$, and different true values of the dependence-related parameter $\Lambda$.

\subsection{Modified Information Criterion (MIC)}
The Modified Information Criterion (MIC) was proposed by \cite{chen2006information}. 
\cite{said2019modified} pointed out that MIC performs better than LRT in detecting the changepoints occurring near the very beginning or at the very end of the data sequence. Here again we assume that the underlying data are independent across time.

The idea of MIC is as follows. Similar to the case of LRT, consider an integer $\tau$ between $1$ (included) to $T$. If a change occurs at any $\tau$, we reject our null hypothesis $H_0$, i.e., no changepoint. Under $H_0$, we have $\tau=T$, and then, the MIC is defined as
\begin{equation}
\nonumber \textrm{MIC}(T)\;=\;-2\log L_{H_0}(\widehat{\Lambda})\;+\;dim(\Lambda)\log(T),
\label{2.12}
\end{equation}
where $\widehat{\Lambda}$ is the solution of \eqref{score_eq_H0}. For the BHR distribution, the dimension of the parameter space is dim($\Lambda$) = 1. 
For $1\leq\,\tau < T$, the MIC is defined as
\begin{equation}
\textrm{MIC}(\tau)\;=\;-2\log L_{H_A}(\widehat{\Lambda}_1,\;\widehat{\Lambda}_T)\;+\;\left[2\;\textrm{dim}(\Lambda_1)\;+\;\left(\frac{2\tau}{T}-1\right)^2\right]\log(T);
\label{eq:mic_tau}
\end{equation}
here, for our model, $\textrm{dim}(\Lambda_1) = 1$.
If $\textrm{MIC}(T) \geq \min_{1\leq \tau<T} \textrm{MIC}(\tau)$, we select the model with a changepoint. Then, the estimated changepoint $\widehat{\tau}$ satisfies the equality $\textrm{MIC}(\widehat{\tau})=\min_{1\leq \tau < T} \textrm{MIC}(\tau)$. 
\cite{chen2006information} also suggested a test statistics $S_T$ based on $\textrm{MIC}(T)$ and $\textrm{MIC}(\tau)$ to detect one changepoint as follows
\begin{equation}
\nonumber S_T=\textrm{MIC}(T)\;-\;\min_{1\leq \tau<T}\;\textrm{MIC}(\tau)\;+\;dim(\Lambda)\log(T).
\label{2.18}
\end{equation} 
For our model, dim$(\Lambda)$ = 1, and hence, $S_T$ can be written for our case as
\begin{equation}
\nonumber S_T=-2\log L_{H_0}(\widehat{\Lambda})-\min_{1\leq \tau<T}\left[-2\log L_{H_A}(\widehat{\Lambda}_1,\widehat{\Lambda}_T)+\left\lbrace \left(\frac{2\tau}{T}-1\right)^2 - 1 \right\rbrace \log (T)\right].
\end{equation}
If changepoints occur at the very beginning or the end, then we do have not sufficient data to obtain the MLEs $\widehat{\Lambda}_1$ or $\widehat{\Lambda}_T$. In these cases,
\cite{ngunkeng2014information} suggested a trimmed version of $S_T$ as
\begin{equation} 
\begin{aligned}
S'_T=-2\log L_{H_0}(\widehat{\Lambda})-\min_{\tau_0 < \tau < T - \tau_0}\left[-2\log L_{H_A}(\widehat{\Lambda}_1,\widehat{\Lambda}_T)+\left\lbrace \left(\frac{2\tau}{T}-1\right)^2 - 1 \right\rbrace \log (T)\right].
\end{aligned}
\label{modified_St}
\end{equation}
where $\tau_0=2 \lfloor \log(T) \rfloor$. Applications of $S'_T$ can be found in \cite{tian2022changepoint}. Once we reject our null hypothesis, the estimated changepoint is
\begin{equation}
\widehat{\tau}_{\textrm{MIC}}\;=\;\argmin_{\tau_0<\tau<T-\tau_0} \;\textrm{MIC}(\tau).
\label{tauhat_MIC}
\end{equation}

We reject our null hypothesis if $S'_T > c_{\alpha,\;T}$, where $c_{\alpha,\;T}$ is the critical value for a given level of significance $\alpha$. We numerically obtain $c_{\alpha,\;T}$ for different choices of $T$, $\alpha$, and true dependence-related parameter $\Lambda$, and we tabulate them in Section \ref{sec:Simulation}.

\section{Simulation Study}
\label{sec:Simulation}
Here we discuss obtaining critical values for the LRT statistic $Z'_T$ in \eqref{modified_Zt} and the MIC statistic $S'_T$ in \eqref{modified_St} and compare their performances in terms of power. We derive the critical values and power numerically due to their intractable analytic expressions.

\subsection{Critical Values} \label{subsec:critical_vals}
We numerically evaluate the critical values $c_{\alpha, T}$ for a few specific choices of the true parameter values $\Lambda$ of the BHR distribution, levels of significance $\alpha$, and sample sizes $T$, to illustrate the procedure for obtaining the critical values under a general set-up as well as for studying their patterns. We consider three choices of $\Lambda \in \{0.5, 2, 4\}$ under the null distribution, four choices of $T \in \{ 50,100,150, 200 \}$, and $\alpha \in \{0.01, 0.05, 0.1\}$.

For a choice of the true value of $\Lambda$ and $T$, we first obtain $B=10^4$ samples each comprising of $T$ IID observations from the BHR distribution with dependence-related parameter $\Lambda$, say $\mathcal{R}^{(b)} = \{ \widetilde{\bm{R}}^{(b)}_1, \ldots, \widetilde{\bm{R}}^{(b)}_T \}$ for $b=1, \ldots, B$. Then, we obtain $\widehat{\Lambda}^{(b)}$, the MLE of $\Lambda$ based on $\mathcal{R}^{(b)}$, following \eqref{score_eq_H0}, and then calculate $\log L_{H_0}(\widehat{\Lambda}^{(b)})$ based on $\mathcal{R}^{(b)}$, following \eqref{log_likelihood_H0}. Further, for each $\tau$ between $\tau_0=2 \lfloor \log(T) \rfloor$ and $T-\tau_0$ (excluding the endpoints), we divide the sample $\mathcal{R}^{(b)}$ into two parts $\mathcal{R}^{(b,1)} = \{\widetilde{\bm{R}}^{(b)}_1, \ldots, \widetilde{\bm{R}}_{\tau}^{(b)}\}$ and $\mathcal{R}^{(b,T)} = \{\widetilde{\bm{R}}_{\tau+1}^{(b)}, \ldots, \widetilde{\bm{R}}^{(b)}_T\}$. We then assume that the observations in $\mathcal{R}^{(b,1)}$ and $\mathcal{R}^{(b,T)}$ follow two different BHR distributions with parameters $\Lambda_1$ and $\Lambda_T$, respectively, and obtain the MLEs $\widehat{\Lambda}^{(b)}_1$ and $\widehat{\Lambda}^{(b)}_T$ following \eqref{score_eq_HA}. Based on $\mathcal{R}^{(b,1)}$, $\mathcal{R}^{(b,T)}$, $\widehat{\Lambda}^{(b)}_1$, and $\widehat{\Lambda}^{(b)}_T$, we calculate $\log L_{H_A}(\widehat{\Lambda}^{(b)}_1,\widehat{\Lambda}^{(b)}_T)$ following \eqref{log_likelihood_HA}. We repeat the above procedure for each $\tau_0 < \tau < T - \tau_0$ and calculate $Z'_T$ following \eqref{modified_Zt}; we call it $Z^{'(b)}_T$. We repeat the whole procedure for each $b=1, \ldots, B$ and obtain $\mathcal{Z}_T = \{Z^{'(1)}_T, \ldots, Z^{'(B)}_T \}$. Similar to LRT statistics, based on $\log L_{H_0}(\widehat{\Lambda}^{(b)})$ and $\log L_{H_A}(\widehat{\Lambda}^{(b)}_1,\widehat{\Lambda}^{(b)}_T)$ for all $\tau_0 < \tau < T - \tau_0$, we calculate $S'_T$ following \eqref{modified_St}; we call it $S^{'(b)}_T$. We repeat the whole procedure for each $b=1, \ldots, B$ and obtain $\mathcal{S}_T = \{S^{'(1)}_T, \ldots, S^{'(B)}_T \}$. Finally, the critical values $c_{\alpha, T}$ for LRT and MIC are obtained by the $100(1-\alpha)$-th percentiles of $\mathcal{Z}_T$ and $\mathcal{S}_T$, respectively. Due to the inherent randomness of the above-explained sampling-based procedure, we also calculate the standard error of $c_{\alpha, T}$ for LRT and MIC using a straightforward nonparametric bootstrap procedure from $\mathcal{Z}_T$ and $\mathcal{S}_T$, as otherwise repeating the parametric bootstrap procedure several times would be computationally challenging. The critical values for both LRT and MIC and their (Monte Carlo) standard errors (S.E.) are presented in Table \ref{table1}. For smaller $\alpha$, $c_{\alpha, T}$ are naturally higher. For a specific $\Lambda$ and $\alpha$, $c_{\alpha, T}$ are generally higher as $T$ increases; however, the underlying high S.E.s indicate that such differences are generally insignificant. The S.E.s of $c_{\alpha, T}$ are high for small $\alpha$ for all $\Lambda$ and $T$ for both LRT and MIC. By construction, the $c_{\alpha, T}$ values are smaller for MIC than that for LRT for any $\Lambda$, $T$, and $\alpha$.


\begin{center}
\begin{table}[h]\small
\renewcommand{\arraystretch}{0.8} 
\caption{Critical values of LRT and MIC for different levels of significance $\alpha$ and different sample sizes $T$.}
\resizebox{\textwidth}{!}{
\label{table1}
\begin{tabular}{ccccccccccccc}
\toprule 
LRT \\
\midrule
$T$ &
$\Lambda$ & & \textbf{$\alpha\,=\,0.01$} & \textbf{$\alpha\,=\,0.05$} & \textbf{$\alpha\,=\,0.1$} & $T$ & $\Lambda$ & & $\alpha\,=\,0.01$ & $\alpha\,=\,0.05$ & $\alpha\,=\,0.1$
\tabularnewline
\midrule
50 & 0.5 & Cutoff & 7.917 & 4.532 & 3.285 & 100 & 0.5 & Cutoff & 7.996 & 4.860 & 3.535 \\
\vspace{-1.5mm}
\tabularnewline
 &  & S.E. & 0.216 & 0.084 & 0.049 & & & S.E. & 0.236 & 0.084 & 0.052
 \tabularnewline
\midrule 
 & 2 & Cutoff & 8.737 & 5.617 & 4.335 &  & 2 & Cutoff & 9.529 & 5.978 & 4.545 \\
 \vspace{-1.5mm}
\tabularnewline
 &  & S.E. & 0.226 & 0.085 & 0.062 & &  & S.E. & 0.177 & 0.097 &
0.078 
\tabularnewline
\midrule 
 & 4 & Cutoff & 8.580 & 5.589 & 4.238 &  & 4 & Cutoff & 9.225 & 5.976 & 4.612 \\
 \vspace{-1.5mm}
\tabularnewline
 &  & S.E. & 0.181 & 0.085 & 0.059 & &  & S.E. & 0.231 & 0.089 & 0.053
\tabularnewline
\midrule 
150 & 0.5 & Cutoff & 7.966 & 4.948 & 3.654 & 200 & 0.5 & Cutoff & 8.036 & 4.968 & 3.698 \\
\vspace{-1.5mm}
\tabularnewline
 &  & S.E. & 0.185 & 0.086 & 0.053 & & & S.E. & 0.210 & 0.087 & 0.055
\tabularnewline
\midrule 
 & 2 & Cutoff & 9.405 & 6.178 & 4.738 &  & 2 & Cutoff & 9.676 & 6.155 & 4.807 \\
 \vspace{-1.5mm}
\tabularnewline
 &  & S.E. & 0.278 & 0.087 & 0.069 & &  & S.E. & 0.207 & 0.087 &
0.059 
\tabularnewline
\midrule 
 & 4 & Cutoff & 9.369 & 6.112 & 4.692 &  & 4 & Cutoff & 9.588 & 6.263 & 4.786 \\
 \vspace{-1.5mm}
\tabularnewline
 &  & S.E. & 0.171 & 0.095 & 0.067 & &  & S.E. & 0.238 & 0.092 &
 0.079 
\tabularnewline
\toprule 
MIC \\ 
\midrule
$T$ &
$\Lambda$ & & \textbf{$\alpha\,=\,0.01$} & \textbf{$\alpha\,=\,0.05$} & \textbf{$\alpha\,=\,0.1$} & $T$ & $\Lambda$ & & $\alpha\,=\,0.01$ & $\alpha\,=\,0.05$ & $\alpha\,=\,0.1$
\tabularnewline
\midrule
50 & 0.5 & Cutoff & 6.723 & 3.553 & 2.406 & 100 & 0.5 & Cutoff & 6.181 & 3.392 & 2.238 \\
\vspace{-1.5mm}
\tabularnewline
 &  & S.E. & 0.236 & 0.081 & 0.052 & & & S.E. & 0.191 & 0.099 & 0.044
 \tabularnewline
\midrule 
  & 2 & Cutoff & 7.510 & 4.540 & 3.292 &  & 2 & Cutoff & 7.600 & 4.200 & 2.957 \\
  \vspace{-1.5mm}
\tabularnewline
 &  & S.E. & 0.213 & 0.087 & 0.056 & & & S.E. & 0.257 & 0.082 & 0.053
 \tabularnewline
\midrule
 & 4 & Cutoff & 7.385 & 4.559 & 3.273 &  & 4 & Cutoff & 7.434 & 4.161 & 2.953 \\
 \vspace{-1.5mm}
\tabularnewline
 &  & S.E. & 0.165 & 0.081 & 0.051 & & & S.E. & 0.167 & 0.061 &
0.056 
\tabularnewline
\midrule
150 & 0.5 & Cutoff & 6.016 & 3.271 & 2.179 & 200 & 0.5 & Cutoff & 5.440 & 3.008 & 2.017 \\
\vspace{-1.5mm}
\tabularnewline
 &  & S.E. & 0.193 & 0.064 & 0.045 & & & S.E. & 0.205 & 0.068 & 0.039
 \tabularnewline
\midrule 
  & 2 & Cutoff & 6.896 & 3.981 & 2.806 &  & 2 & Cutoff & 7.031 & 3.747 & 2.508 \\
  \vspace{-1.5mm}
\tabularnewline
 &  & S.E. & 0.252 & 0.073 & 0.051 & & & S.E. & 0.249 & 0.079 & 0.052  
 \tabularnewline
\midrule 
  & 4 & Cutoff & 7.164 & 4.029 & 2.749 &  & 4 & Cutoff & 6.778 & 3.743 & 2.535 \\
  \vspace{-1.5mm}
\tabularnewline
 &  & S.E. & 0.200 & 0.083 & 0.060 & & & S.E. & 0.228 & 0.069 & 0.054  
 \tabularnewline
\bottomrule
\end{tabular}}
\vspace{0.2cm}
\end{table}
\end{center}

\vspace{-15mm}
\subsection{Power Comparison} \label{subsec:power}
In this subsection, we compare the power of LRT and MIC numerically under different scenarios of the alternative hypothesis $H_A$, i.e., we fix the changepoint $\tau$, the values of the dependence-related parameter of the BHR distribution before and after the changepoint, i.e., $\Lambda_1$ and $\Lambda_T$. We choose the sample sizes $T \in \{50, 200\}$ and the levels of significance $\alpha \in \{0.01, 0.05, 0.1\}$. Then, for each $T$, we choose two different values of the changepoint $\tau = \lfloor \beta T \rfloor$, where $\beta \in \{0.25, 0.5 \}$. In each case, we consider the possible values of the dependence-related parameter before and after the changepoint as $\Lambda_1 \in \{0.5, 2, 4\}$ and $\Lambda_T \in \{0.5, 1, 1.5, \ldots, 5\}$, respectively. Under $H_0$, $\tau=T$ and the value of the dependence-related parameter under $H_0$ is the same as $\Lambda_1$. Hence, for the above choices, the critical values of LRT and MIC are obtained as in Section \ref{subsec:critical_vals}.

For a choice of the true values of $\Lambda_1$, $\Lambda_T$, $\tau$, and $T$, we first obtain $B=10^4$ samples each comprising of $\tau$ IID observations from the BHR distribution with dependence-related parameter $\Lambda_1$, say $\mathcal{R}^{(b,1)} = \{\widetilde{\bm{R}}^{(b)}_1, \ldots, \widetilde{\bm{R}}_{\tau}^{(b)}\}$ and another $T-\tau$ IID observations from the BHR distribution with parameter $\Lambda_T$, say $\mathcal{R}^{(b,T)} = \{\widetilde{\bm{R}}_{\tau+1}^{(b)}, \ldots, \widetilde{\bm{R}}^{(b)}_T\}$, for $b=1, \ldots, B$. For each $b$, we thus obtain a combined sample of size $T$ given by $\mathcal{R}^{(b)} = \{\widetilde{\bm{R}}^{(b)}_1, \ldots, \widetilde{\bm{R}}^{(b)}_T \}$. Under $H_0$, we assume the observations in $\mathcal{R}^{(b)}$ to be IID following a BHR distribution with a common parameter $\Lambda$. Following \eqref{score_eq_H0}, we obtain the MLE of $\Lambda$ given by $\widehat{\Lambda}^{(b)}$ and then calculate $\log L_{H_0}(\widehat{\Lambda}^{(b)})$ based on $\mathcal{R}^{(b)}$, following \eqref{log_likelihood_H0}. Subsequently, under $H_A$, we obtain the MLEs $\widehat{\Lambda}^{(b)}_1$ and $\widehat{\Lambda}^{(b)}_T$ following \eqref{score_eq_HA}. Based on $\mathcal{R}^{(b,1)}$, $\mathcal{R}^{(b,T)}$, $\widehat{\Lambda}^{(b)}_1$, and $\widehat{\Lambda}^{(b)}_T$, we calculate $\log L_{H_A}(\widehat{\Lambda}^{(b)}_1,\widehat{\Lambda}^{(b)}_T)$ following \eqref{log_likelihood_HA}. We repeat the above procedure for each $\tau_0 < \tau < T - \tau_0$ and calculate $Z'_T$ following \eqref{modified_Zt}; we call it $Z^{'(b)}_T$. We repeat the whole procedure for each $b=1, \ldots, B$ and obtain $\mathcal{Z}_T = \{Z^{'(1)}_T, \ldots, Z^{'(B)}_T \}$. Similar to LRT statistics, based on $\log L_{H_0}(\widehat{\Lambda}^{(b)})$ and $\log L_{H_A}(\widehat{\Lambda}^{(b)}_1,\widehat{\Lambda}^{(b)}_T)$ for all $\tau_0 < \tau < T - \tau_0$, we calculate $S'_T$ following \eqref{modified_St}; we call it $S^{'(b)}_T$. We repeat the whole procedure for each $b=1, \ldots, B$ and obtain $\mathcal{S}_T = \{S^{'(1)}_T, \ldots, S^{'(B)}_T \}$. Despite having a true value of $\tau$ under $H_A$, both LRT and MIC should be calculated based on maximizing the necessary terms in \eqref{modified_Zt} and \eqref{modified_St} over $\tau_0 < \tau < T - \tau_0$. The critical values are obtained apriori following Section \ref{subsec:critical_vals} and suppose we call the critical values $c_{\alpha, T}$ for LRT and MIC by $c^{\textrm{LRT}}_{\alpha, T}$ and $c^{\textrm{MIC}}_{\alpha, T}$, respectively.
We thus can approximate their powers by 
\begin{equation}
P_{\textrm{LRT}}=\frac{1}{B}\sum_{b=1}^B \textrm{I}\left(Z_T^{'{(b)}}\geq c^{\textrm{LRT}}_{\alpha, T} \right), \quad P_{\textrm{MIC}}=\frac{1}{B}\sum_{b=1}^B \textrm{I}\left(S_T^{'(b)}\geq c^{\textrm{MIC}}_{\alpha, T}\right),
\label{eq:power} 
\end{equation}
where I($\cdot$) is an indicator function.

\begin{figure}[!t]
\centering
\includegraphics[height = 0.25\linewidth]{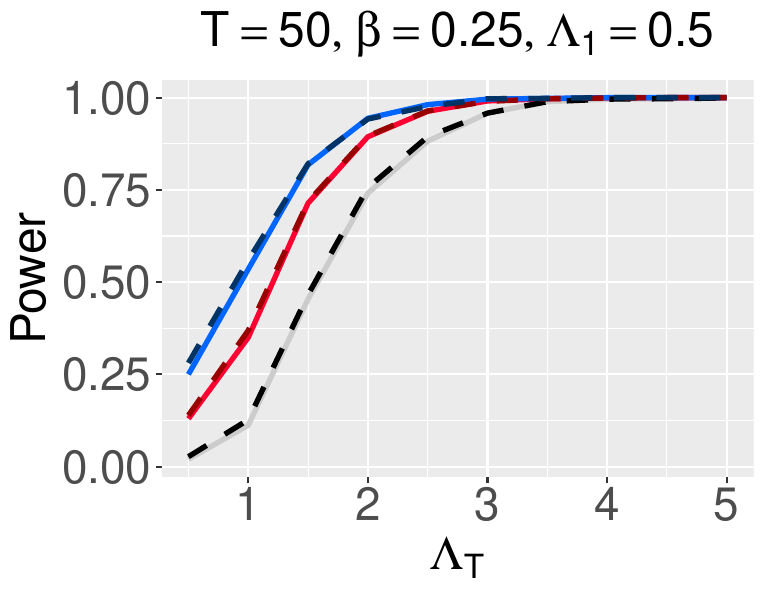}
\includegraphics[height = 0.25\linewidth]{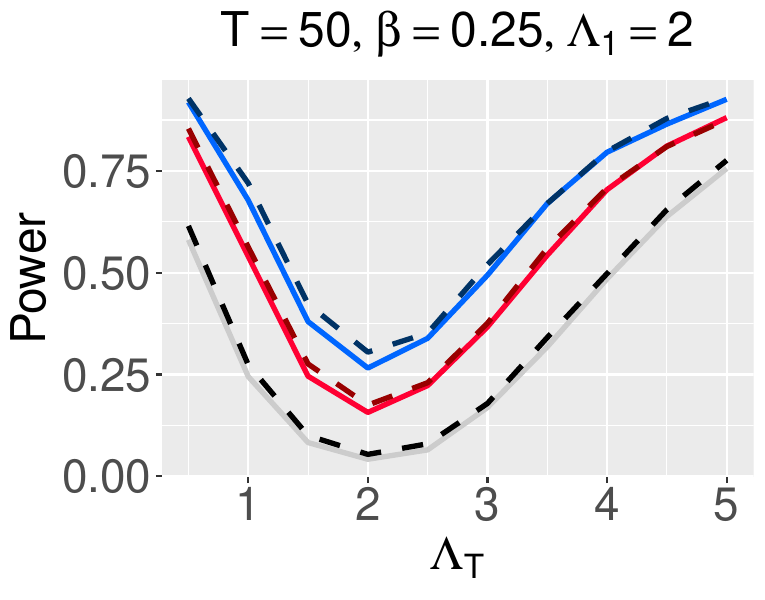}
\includegraphics[height = 0.25\linewidth]{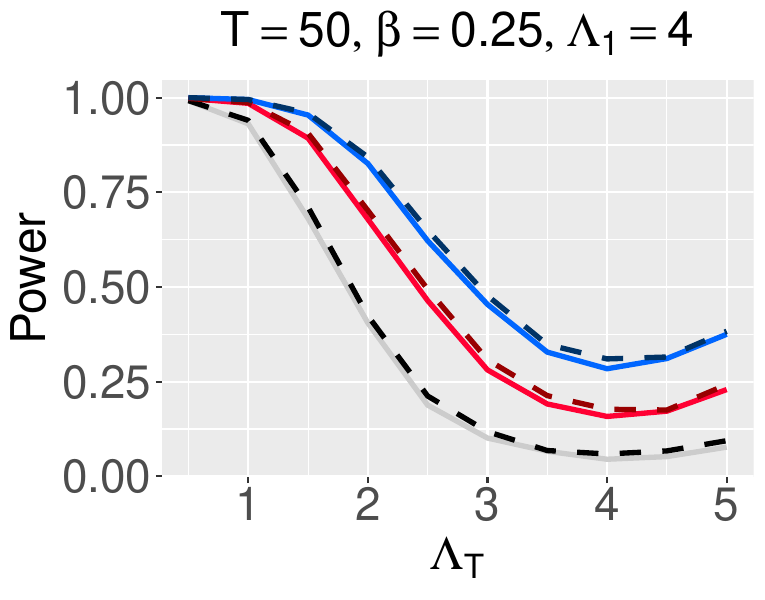}\\
\includegraphics[height = 0.25\linewidth]{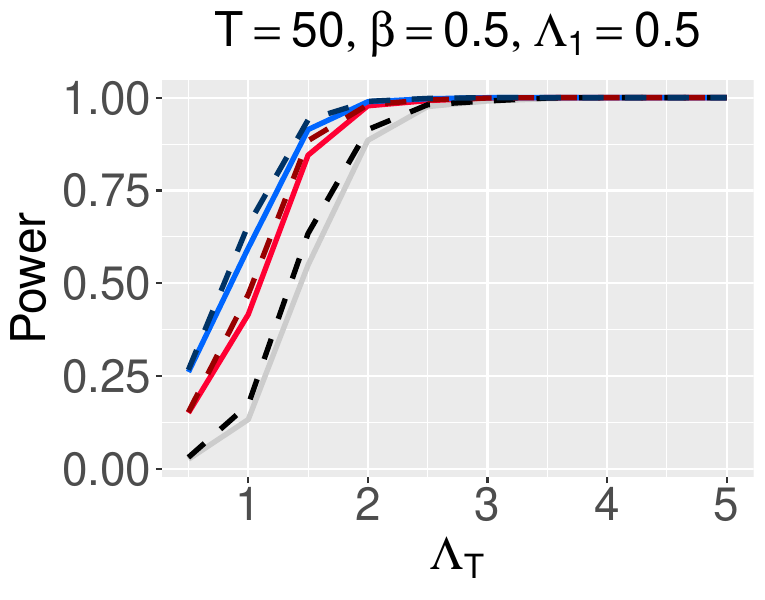}
\includegraphics[height = 0.25\linewidth]{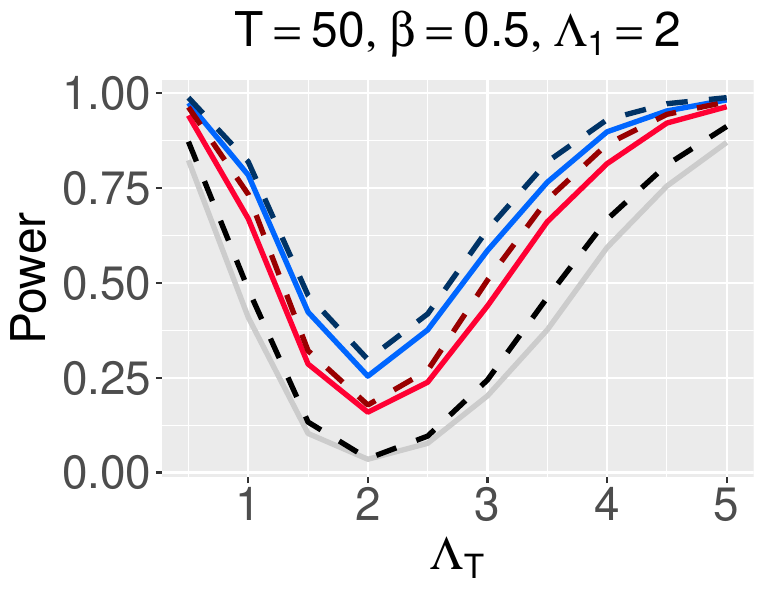}
\includegraphics[height = 0.25\linewidth]{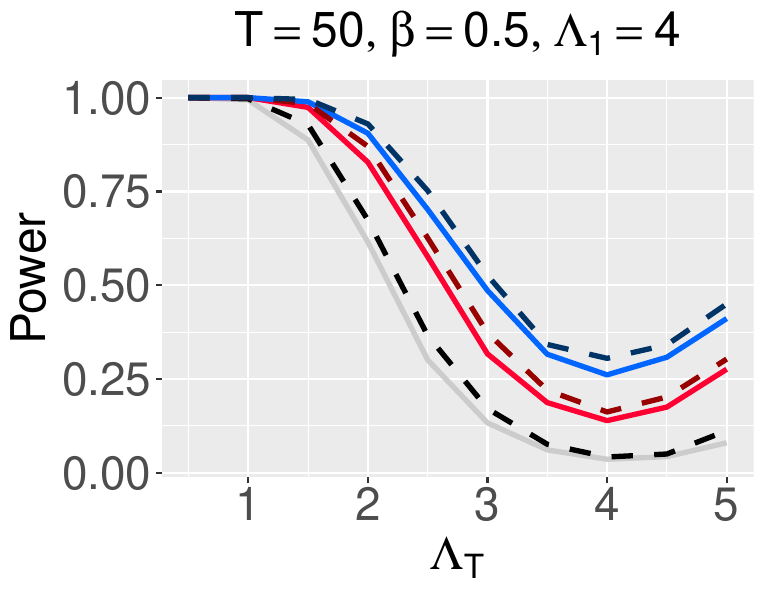} \\
\includegraphics[height = 0.25\linewidth]{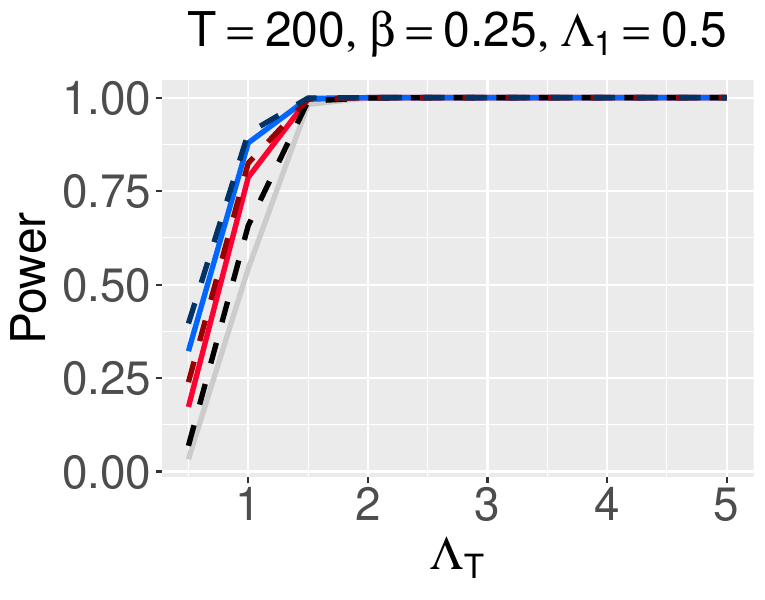}
\includegraphics[height = 0.25\linewidth]{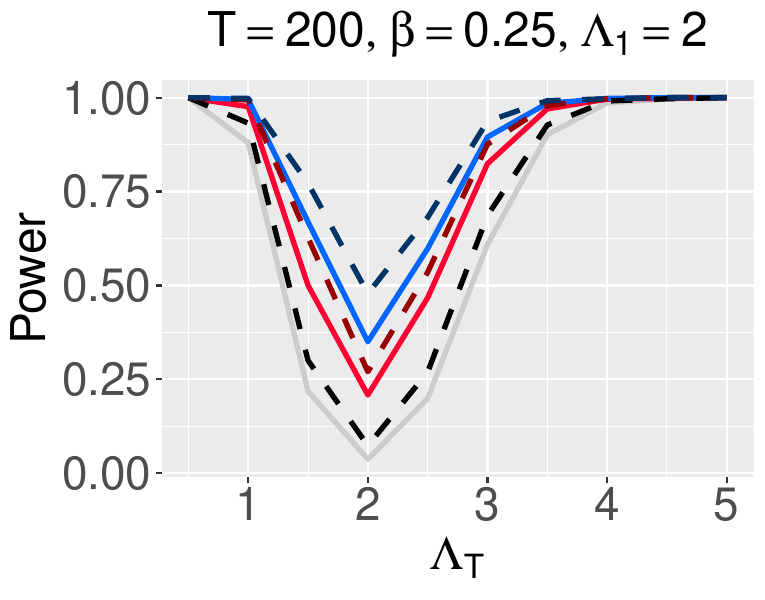}
\includegraphics[height = 0.25\linewidth]{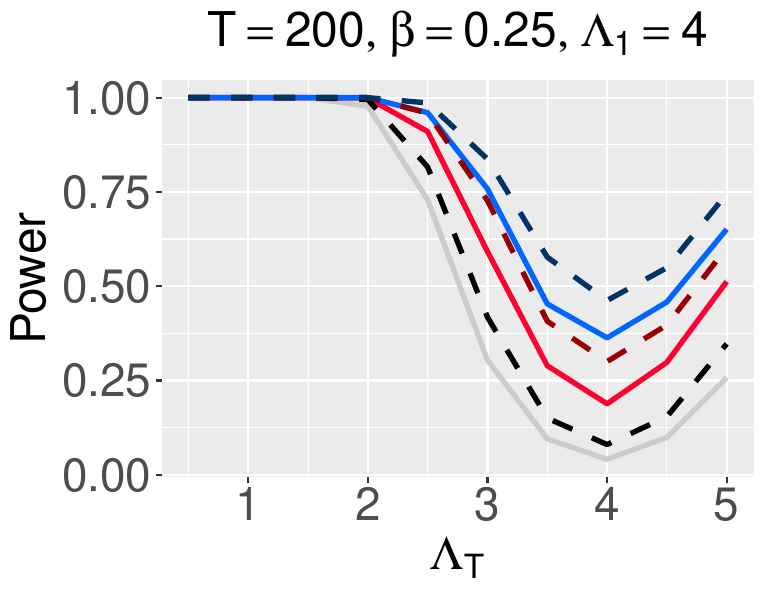}\\
\includegraphics[height = 0.25\linewidth]{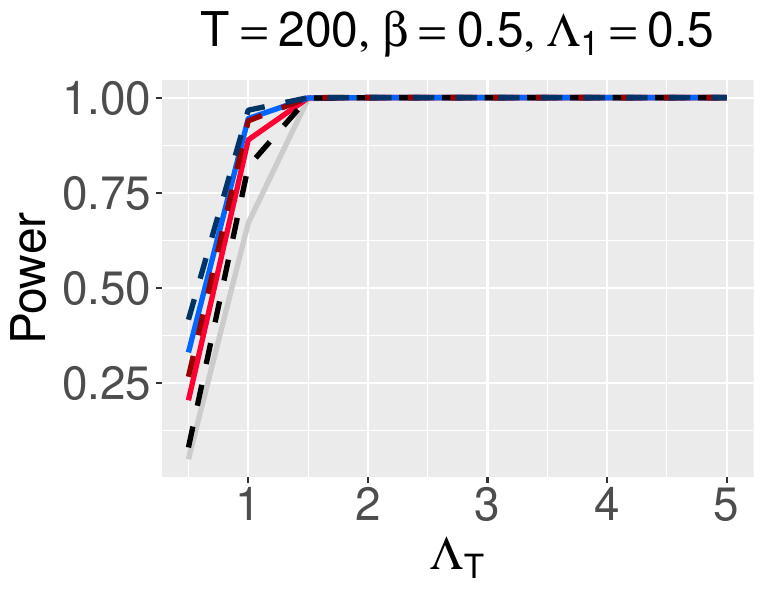}
\includegraphics[height = 0.25\linewidth]{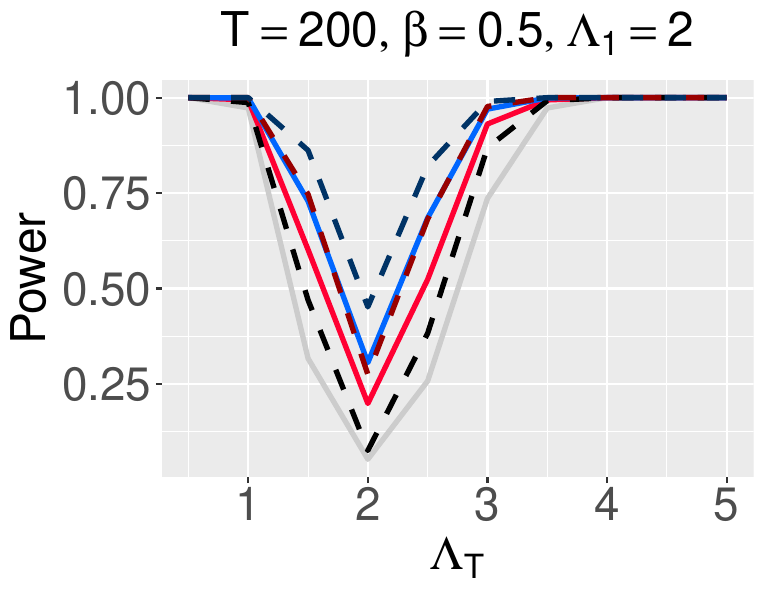}
\includegraphics[height = 0.25\linewidth]{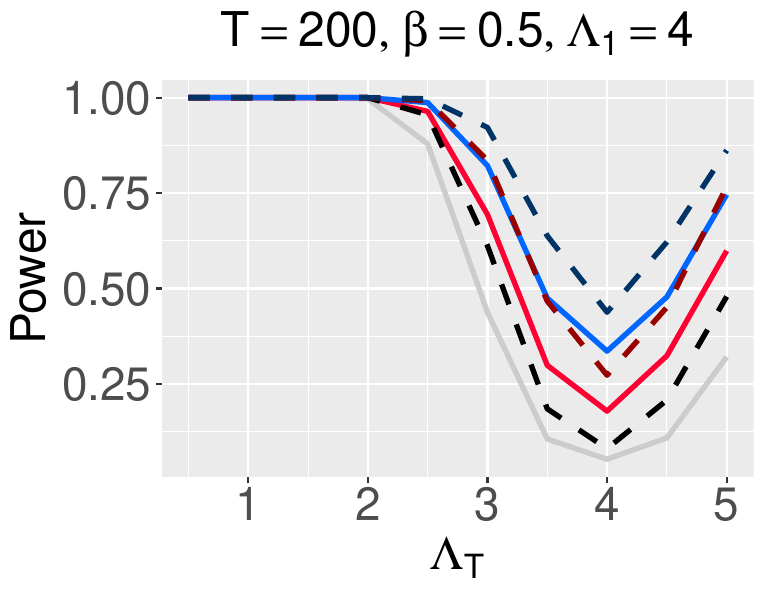}
\vspace{-0.5cm}
\caption{Power curves of LRT and MIC under different choices of $T$, $\tau = \lfloor \beta T\rfloor$, $\Lambda_1$, $\Lambda_T$, and $\alpha$. The curves for LRT are presented by solid lines while the curves for MIC are presented by dashed lines. For each case, the curves from bottom to top, represented in black, red, and blue, are for $\alpha = 0.01, 0.05, 0.1$, respectively. A higher value of power indicates a better testing procedure.}
\label{fig:powerplots}
\end{figure}


We present the power curves in Figure \ref{fig:powerplots}. For both LRT and MIC, the power is higher for larger levels of significance $\alpha$ under all settings; this is obvious and follows directly from \eqref{eq:power}. When $T=50$, $\beta = 0.25$, and $\Lambda_1 = 0.5$, the power curves of LRT and MIC for $\alpha = 0.01$ are close to one only when $\Lambda_T \approx 3.5$. After changing $\beta=0.25$ to $\beta= 0.5$ keeping other parameters fixed, both power curves for $\alpha = 0.01$ are close to one when $\Lambda_T \approx 2.5$. Here, the availability of more observations for estimating $\Lambda_1$ allows higher power even when the difference between $\Lambda_1$ and $\Lambda_T$ remains the same. In general, the power of both LRT and MIC are higher for $\beta=0.5$ than for $\beta=0.25$ while keeping the other parameters fixed. This feature is more prominent when the sample size is small. Keeping other parameters fixed (with respect to the top-left panel), increasing the sample size from $T=50$ to $T=200$, the power increases significantly for both tests; both power curves for $\alpha = 0.01$ are close to one when $\Lambda_T \approx 1.5$. A similar increasing pattern in power curves is visible for other values of $\Lambda_1$ and $\beta$. In all cases, naturally, the power is lowest when $\Lambda_1 = \Lambda_T$. For $T=50$ and $\Lambda_1=2$, we observe that the power curves are asymmetric for $\Lambda_T < \Lambda_1$ and $\Lambda_T > \Lambda_1$. When $\Lambda_T < \Lambda_1$, the power values of LRT and MIC get closer to one faster than that in the case of $\Lambda_T > \Lambda_1$. For $T=200$, the power curves become more symmetric around $\Lambda_1$. However, for this asymmetry changes with the value of $\Lambda_1$; for example, when $\Lambda_1 = 0.5$ and $T=200$, the power curves reach near one when $\Lambda_T - \Lambda_1 \approx 1$, while for $\Lambda_1 = 4$ and $T=200$, all the power curves become close to one only when $\Lambda_1 - \Lambda_T \approx 2$. In general, MIC appears to be more powerful than LRT. This characteristic is more prominent when $T=200$. For both $T=50$ and $T=200$, the differences between the power curves of LRT and MIC are generally more prominent when $\beta=0.5$ compared to the case of $\beta=0.25$.

\subsection{Consistency of the Estimator of Changepoint} \label{subsec:consistency}
There is a vast literature on the asymptotic theoretical guarantees of both LRT \citep{csorgo1997limit} and MIC \citep{chen2006information}. They hold in general under certain regularity conditions. However, despite the availability of asymptotic results, an analytical exposition of finite sample properties is often intricate because of the non-existence of any closed-form expression of the estimators of changepoint $\widehat{\tau}_{\textrm{LRT}}$ in \eqref{tauhat_LRT} and $\widehat{\tau}_{\textrm{MIC}}$ in \eqref{tauhat_MIC}. As a result, in this subsection, we numerically calculate the probabilities of the estimators $\widehat{\tau}_{\textrm{LRT}}$ and $\widehat{\tau}_{\textrm{MIC}}$ being within a small neighborhood of the true value of the changepoint $\tau$, i.e., $\textrm{P}(\vert \widehat{\tau} - \tau \vert \leq \delta)$ for $\widehat{\tau} = \widehat{\tau}_{\textrm{LRT}}, \widehat{\tau}_{\textrm{MIC}}$ and $\delta=1,2,3$. We calculate this probability under different scenarios of the alternative hypothesis $H_A$, i.e., we fix the changepoint $\tau$, the values of the dependence-related parameter of the BHR distribution before and after the changepoint, i.e., $\Lambda_1$ and $\Lambda_T$. We choose the sample sizes $T \in \{50, 200\}$. 
Then, for each $T$, we choose two different values of the changepoint $\tau = \lfloor \beta T \rfloor$, where $\beta \in \{0.25, 0.5 \}$. In each case, we consider the possible values of the dependence-related parameter before and after the changepoint as $\Lambda_1 \in \{0.5, 1, 1.5, \ldots, 5\}$ and $\Lambda_T \in \{0.5, 2, 4\}$, respectively. We ignore the cases when $\Lambda_1 = \Lambda_T$ as they indicate no changepoint scenarios.

\begin{figure}[!t]
\includegraphics[height = 0.251\linewidth]{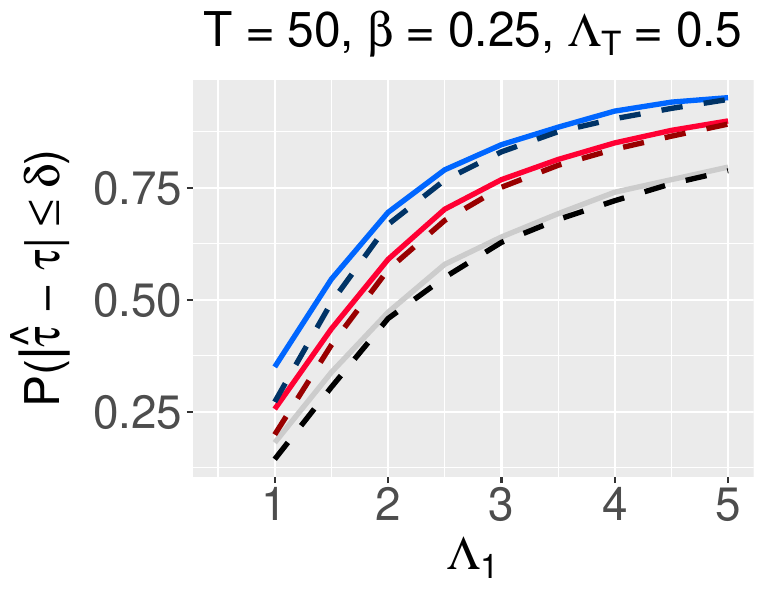}
\includegraphics[height = 0.251\linewidth]{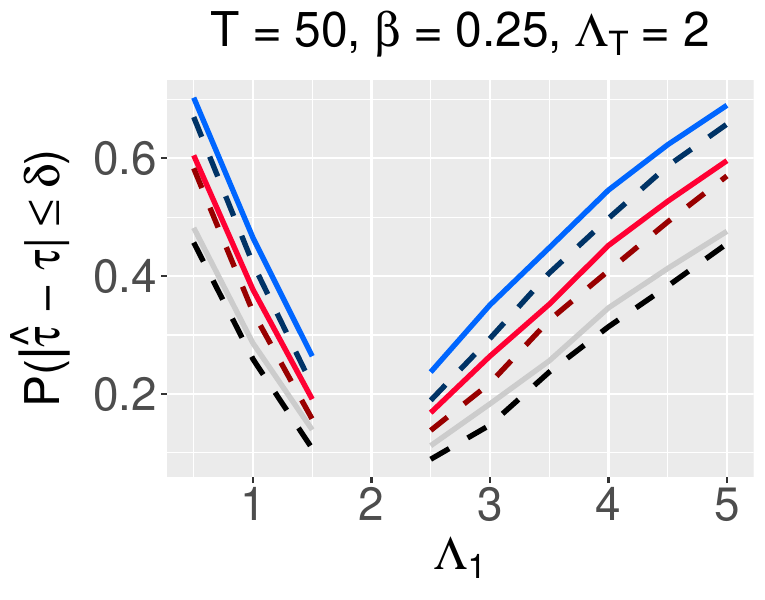}
\includegraphics[height = 0.251\linewidth]{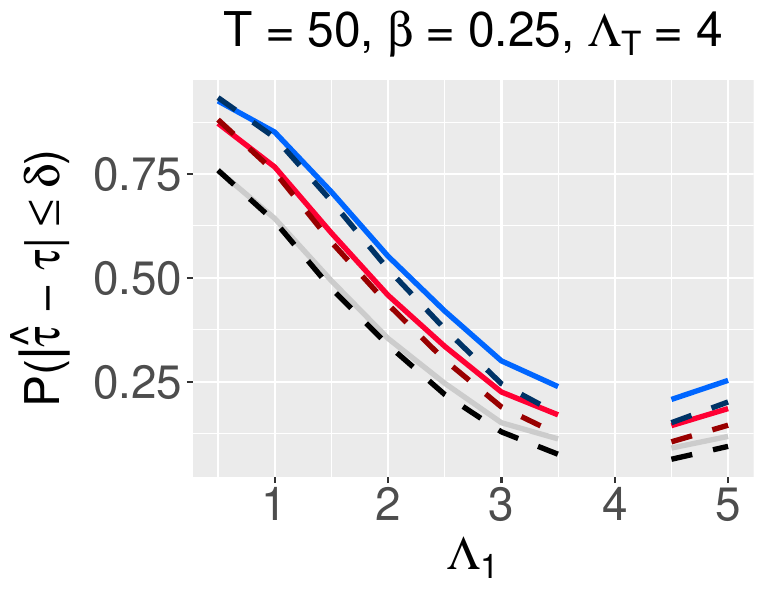}\\
\includegraphics[height = 0.251\linewidth]{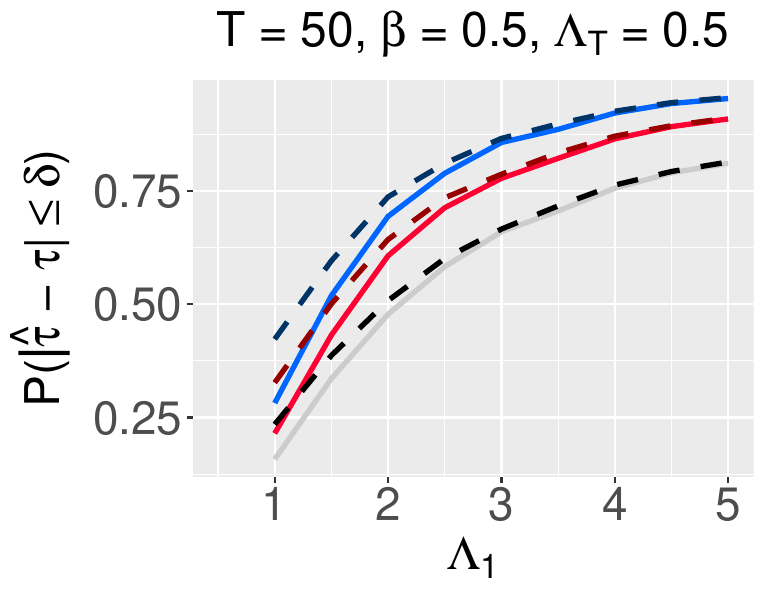}
\includegraphics[height = 0.251\linewidth]{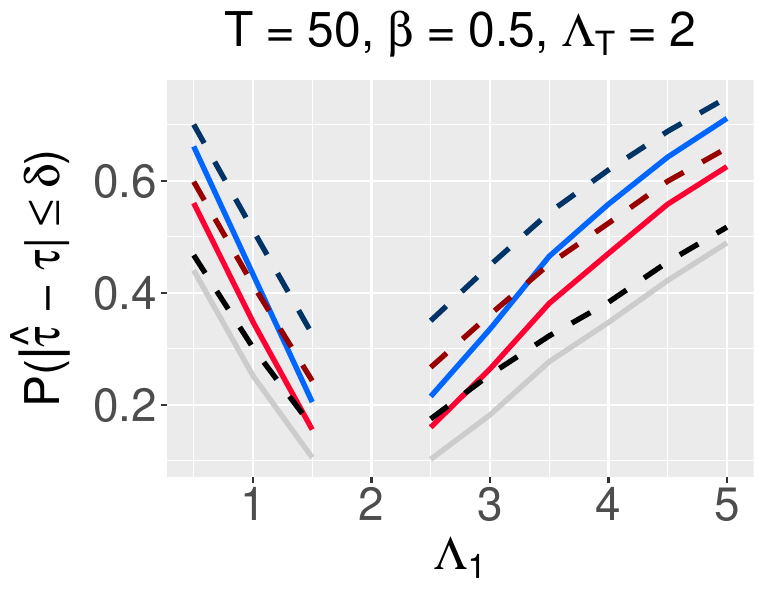}
\includegraphics[height = 0.251\linewidth]{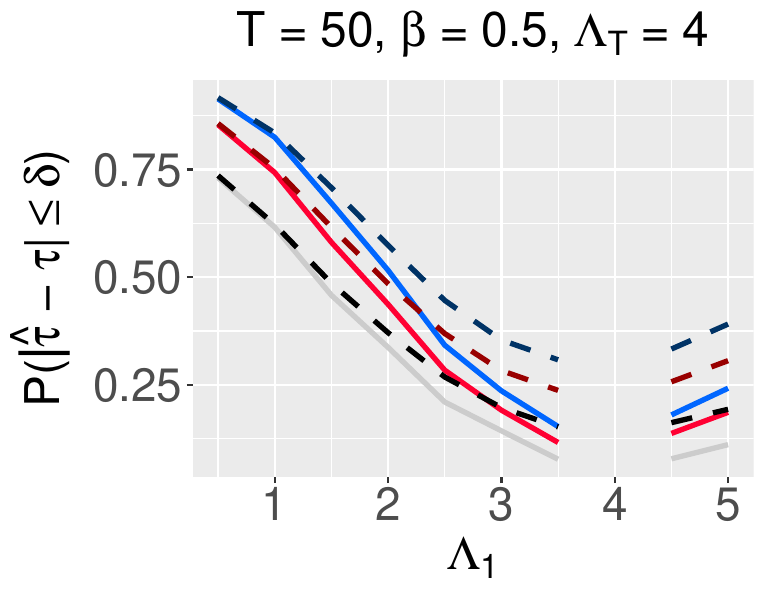} \\
\includegraphics[height = 0.251\linewidth]{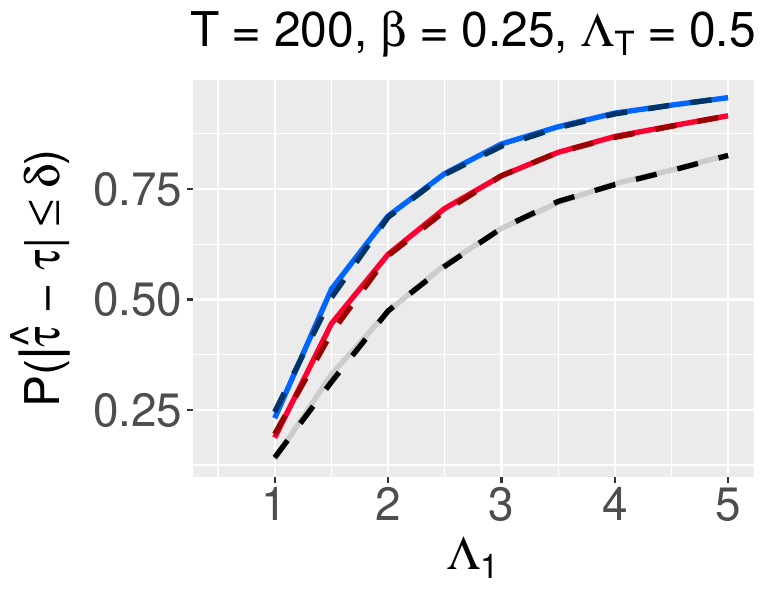}
\includegraphics[height = 0.251\linewidth]{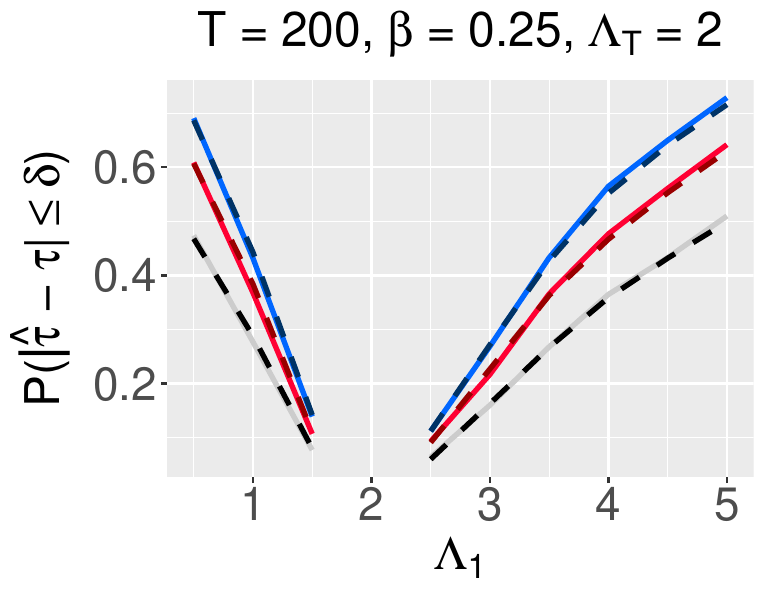}
\includegraphics[height = 0.251\linewidth]{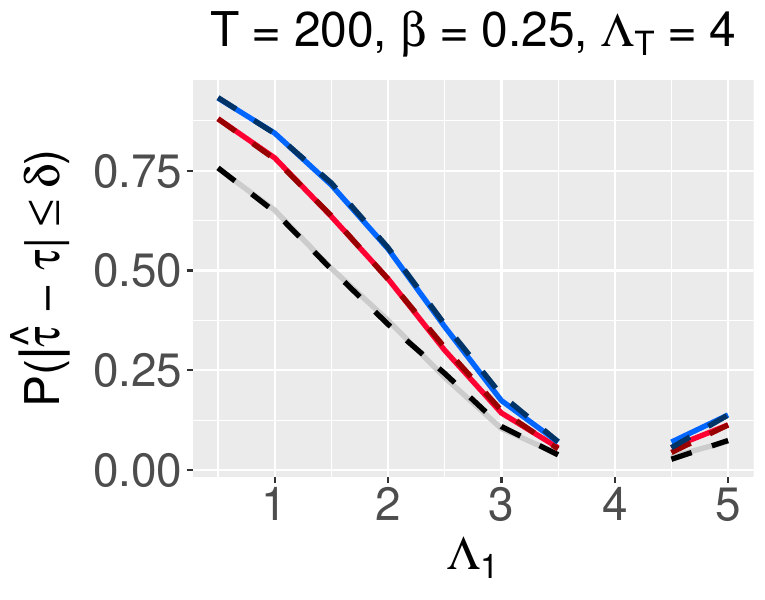}\\
\includegraphics[height = 0.251\linewidth]{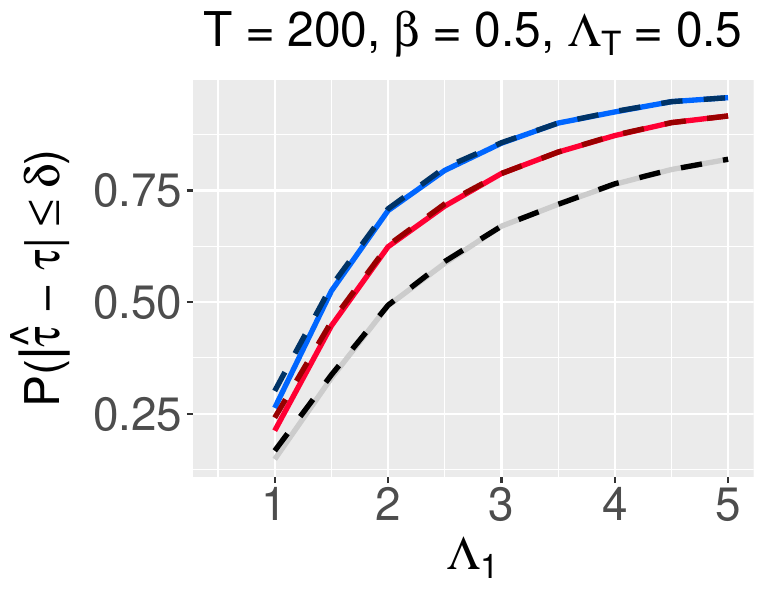}
\includegraphics[height = 0.251\linewidth]{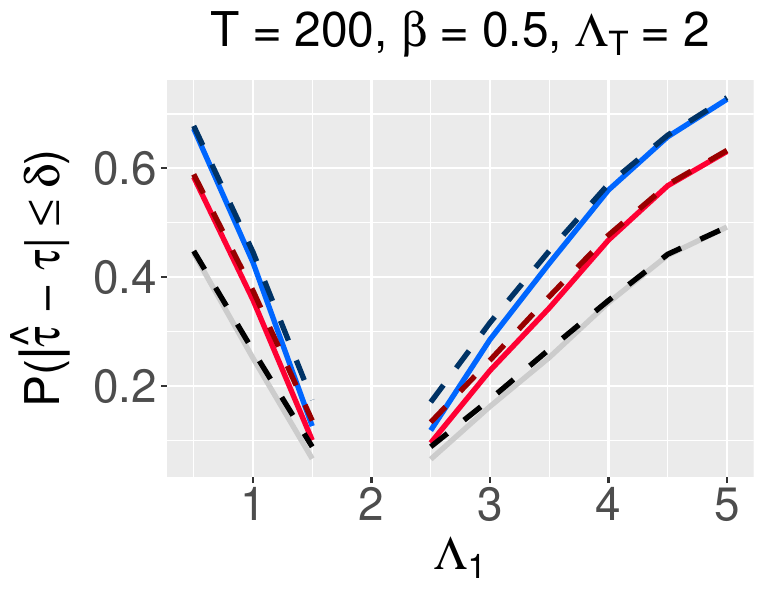}
\includegraphics[height = 0.251\linewidth]{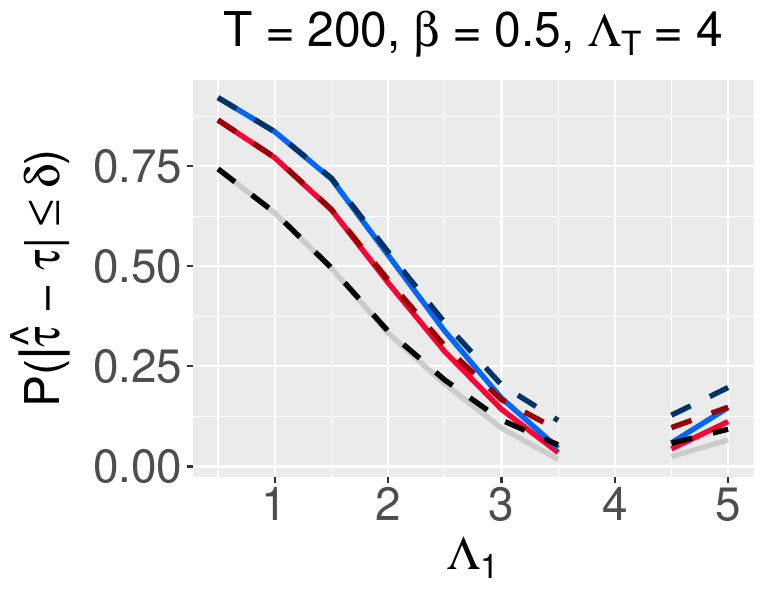}
\vspace{-0.5cm}
\caption{The probabilities $\textrm{P}(\vert \widehat{\tau} - \tau \vert \leq \delta)$ for $\widehat{\tau} = \widehat{\tau}_{\textrm{LRT}}, \widehat{\tau}_{\textrm{MIC}}$ and $\delta \in \{1,2,3\}$ under different choices of $T$, $\tau = \lfloor \beta T\rfloor$, $\Lambda_1$, and $\Lambda_T$. The curves for LRT are presented by solid lines while the curves for MIC are presented by dashed lines. For each case, the curves from bottom to top, represented in black, red, and blue, are for $\delta = 1,2,$ and 3, respectively. A higher value of $\textrm{P}(\vert \widehat{\tau} - \tau \vert \leq \delta)$ indicates a better accuracy in the estimation of a changepoint.} 
\label{fig:consistency}
\end{figure}

We present the curves of $\textrm{P}(\vert \widehat{\tau} - \tau \vert \leq \delta)$ in Figure \ref{fig:consistency}. As $\delta$ increases, $\textrm{P}(\vert \widehat{\tau} - \tau \vert \leq \delta)$ naturally also increases for $\widehat{\tau} = \widehat{\tau}_{\textrm{LRT}}, \widehat{\tau}_{\textrm{MIC}}$ under all settings. When $T=50, \beta=0.25$, and $\Lambda_T = 0.5$ (top-left panel), the values of $\textrm{P}(\vert \widehat{\tau} - \tau \vert \leq \delta)$ are close to each other for LRT and MIC; for small differences between $\Lambda_1$ and $\lambda_T$, LRT performs slightly better than MIC and the difference fades away as the difference between $\Lambda_1$ and $\lambda_T$ increases. Upon changing $\Lambda_T = 0.5$ to $\Lambda_T = 2$, the difference in $\textrm{P}(\vert \widehat{\tau} - \tau \vert \leq \delta)$ between LRT and MIC is more prominent, and LRT generally outperforms MIC. Further, changing $\Lambda_T = 2$ to $\Lambda_T = 4$, we see a less prominent difference in performance between LRT and MIC. When $\beta=0.5$ (second row of Figure \ref{fig:consistency}) instead of $\beta=0.25$ as in the previous cases, we notice an opposite pattern in terms of the performances of LRT and MIC, i.e., MIC performs better than LRT for small differences between $\Lambda_1$ and $\Lambda_T$, and for large values of $\vert \Lambda_1 - \Lambda_T \vert$, both methods perform similarly. When $T=200$, both LRT and MIC perform equally in general, and for $\beta=0.5$, MIC outperforms LRT for small values of $\vert \Lambda_1 - \Lambda_T \vert$. Theoretically, the probability $\textrm{P}(\vert \widehat{\tau} - \tau \vert \leq \delta)$ should converge to one as $T \uparrow \infty$ for any positive $\vert \Lambda_1 - \Lambda_T \vert$; however, we see that for $\vert \Lambda_1 - \Lambda_T \vert =0.5$, the probability is less than 0.5 for all settings we consider. However, corresponding to the bottom-left panel of Figure \ref{fig:consistency}, for example, we observe that the inclusion probability is close to one for $\delta=3$ when $\vert \Lambda_1 - \Lambda_T \vert =4.5$. This observation indicates that the convergence underlying the asymptotic consistency holds only slowly while increasing $T$, and the difference in $\Lambda_1$ and $\Lambda_T$ plays a crucial role here.

We further study (not shown) the bias and mean squared error (MSE) in estimating the changepoint $\tau$ using $\widehat{\tau}_{\textrm{LRT}}$ and $\widehat{\tau}_{\textrm{MIC}}$. We observe that MIC produces significantly smaller values of MSE than LRT, and the difference is more prominent when $T=200$ compared to the case of $T=50$. Both bias and MSE are higher when the difference between $\Lambda_1$ and $\Lambda_T$ is small. We observe that the MSE decreases at a faster rate with $\vert \Lambda_1 - \Lambda_T \vert$ when $\Lambda_1 < \Lambda_T$ compared to the case when $\Lambda_1 > \Lambda_T$. This pattern is consistent across the choices of $\beta$. For $\beta=0.25$, we observe positive biases for both LRT and MIC across all choices of $\Lambda_1$ and $\Lambda_T$. However, for $\beta=0.5$, we observe positive biases when $\Lambda_1 > \Lambda_T$ and negative biases when $\Lambda_1 < \Lambda_T$, irrespective of the method and the value of $T$. The absolute bias based on MIC is slightly higher than that for LRT when $\beta=0.25$, and the reverse is observable when $\beta=0.5$. 


\section{Data Application}
\label{sec:Data_Application}
\subsection{Local Probability Weighted Moments Estimation}
\label{subsec:local_pwm}
\begin{figure}[t]
\centering
\includegraphics[width=0.9\linewidth]{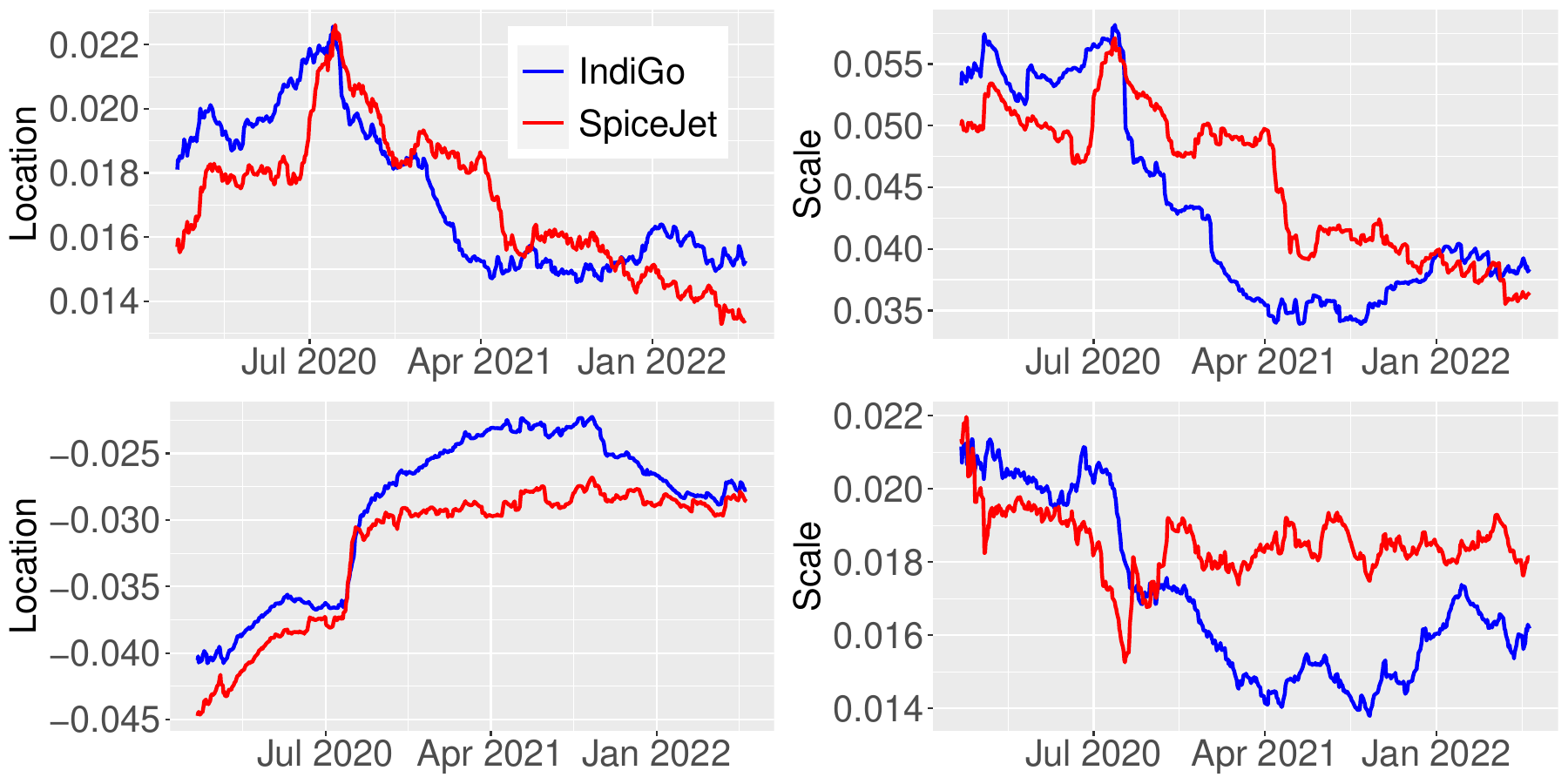}
\caption{Local Probability Weighted Moments estimates of the temporally-varying location (left) and scale (right) parameters of the fitted Gumbel distribution to daily maximum (top) and minimum (bottom) rates of return for IndiGo and SpiceJet airlines.}
\label{fig:local_PWM_estimates} 
\end{figure}

In Figure \ref{fig:returnrate_ts_maxmin}, we observe the nonstationary nature of the daily maximum/minimum rate of return (RoR) for both IndiGo and SpiceJet airlines, and in Section \ref{subsec:data_preprocessing}, we discuss considering the temporally-varying location and scale parameters of the underlying Gumbel distributions. We estimate them using local probability weighted moments estimation and obtain $\{\widehat{\mu}^X_t, \widehat{\sigma}^X_t \}$, the location and scale parameters for IndiGo, and $\{\widehat{\mu}^Y_t, \widehat{\sigma}^Y_t\}$, the location and scale parameters for SpiceJet, respectively, for each of the daily maximum/minimum RoR series. We present the time series of those estimates in Figure \ref{fig:local_PWM_estimates}. For both daily maximum/minimum RoR series, the estimated location and scale profiles are higher during the first half of the COVID-19 period and lower during the second half for both airlines. This observation indicates that the median daily maximum/minimum RoR and variability are higher during the first half of the observation period, i.e., the overall volatility is high. Later, the median daily absolute maximum/minimum RoR and the variability drop. For the daily maximum return of rate, all profiles attain peaks during August 2020; however, we observe a decreasing trend afterward for both profiles of SpiceJet, while the profiles stabilize afterward for IndiGo. This pattern indicates that the median daily maximum RoR are unstable after a volatile first half of the COVID-19 period. Both the estimated location profiles $\{\widehat{\mu}^X_t, \widehat{\mu}^Y_t \}$ remain positive throughout the COVID-19 period. For the daily minimum return of rate, the estimated location profiles gets closer to zero until July 2020 and then remains stable, while the profile for IndiGo remains closer to zero throughout the study period. The scale profiles decrease for both airlines until July 2020 for both airlines and remains stable afterwards. Lower values of the scale profile for IndiGo after July 2020 indicates lower volatility. While our main focus in this paper and the next subsection is on the shifts in the dependence structure, the patterns observable in Figure \ref{fig:local_PWM_estimates} shed light on the marginal behavior of the daily maximum/minimum RoR. 

\subsection{Changepoint Estimation}


Based on the estimated location and scale parameters of the marginal Gumbel distributions in Section \ref{subsec:local_pwm}, we transform the daily maximum/minimum RoR series to standard Gumbel margins according to \eqref{eq:standard_gumbel}. We denote the transformed data for both maximum/minimum RoR series by a generic notation $\mathcal{R} = \{ \widetilde{\bm{R}}_t = (\widetilde{R}^X_t, \widetilde{R}^Y_t), t=1, \ldots, T \}$. We first calculate the LRT and MIC statistics as follows. Under $H_0$, we assume the observations in $\mathcal{R}$ to be IID following a BHR distribution with a common parameter $\Lambda$. Following \eqref{score_eq_H0}, we obtain the MLE of $\Lambda$ given by $\widehat{\Lambda}$ and then calculate $\log L_{H_0}(\widehat{\Lambda})$ based on $\mathcal{R}$, following \eqref{log_likelihood_H0}. Subsequently, under $H_A$, for each possible value of the changepoint $\tau$ such that $\tau_0 < \tau < T - \tau_0$ where $\tau_0 = 2 \lfloor \log(T) \rfloor$, we obtain the MLEs $\widehat{\Lambda}_1$ and $\widehat{\Lambda}_T$ following \eqref{score_eq_HA}, based on $\mathcal{R}^{(1)} = \{ \widetilde{\bm{R}}_t, t=1, \ldots, \tau \}$ and $\mathcal{R}^{(T)} = \{ \widetilde{\bm{R}}_t, t=\tau+1, \ldots, T \}$, respectively. Based on $\mathcal{R}^{(1)}$, $\mathcal{R}^{(T)}$, $\widehat{\Lambda}_1$, and $\widehat{\Lambda}_T$, we calculate $\log L_{H_A}(\widehat{\Lambda}_1,\widehat{\Lambda}_T)$ following \eqref{log_likelihood_HA}. 
Accordingly, we calculate $Z'_T$ following \eqref{modified_Zt}; we call it $Z^{'(obs)}_T$. Similarly, we calculate $S'_T$ following \eqref{modified_St}; we call it $S^{'(obs)}_T$.

Here, we deal with a real dataset, and thus, the true parameter values of the underlying distribution are unknown. Hence, for determining the critical values for both tests and the corresponding $p$-values, we use parametric bootstrapping. For each $b=1, \ldots, B$, where we choose $B=2000$, we draw an IID sample of size $T$, say $\mathcal{R}^{(b)}$, from the BHR distribution with parameter $\widehat{\Lambda}$. We repeat the same procedure of obtaining $Z'_T$ and $S'_T$ based on $\mathcal{R}^{(b)}$ and suppose we call them by $Z^{'(b)}_T$ and $S^{'(b)}_T$, respectively. We repeat the whole procedure for each $b=1, \ldots, B$ and obtain $\mathcal{Z}_T = \{Z^{'(1)}_T, \ldots, Z^{'(B)}_T \}$ and $\mathcal{S}_T = \{S^{'(1)}_T, \ldots, S^{'(B)}_T \}$. We calculate the critical values based on the $100(1-\alpha)$-th percentiles of $\mathcal{Z}_T$ and $\mathcal{S}_T$, respectively. The corresponding $p$-values are obtained by 
\begin{equation}
p_{\textrm{LRT}}=\frac{1}{B}\sum_{b=1}^B \textrm{I}\left(Z_T^{'{(b)}}\geq Z^{'(obs)}_T \right), \quad p_{\textrm{MIC}}=\frac{1}{B}\sum_{b=1}^B \textrm{I}\left(S_T^{'(b)}\geq S^{'(obs)}_T \right).
\label{eq:pval} 
\end{equation}

\begin{figure}[t]
    \centering
    \includegraphics[width=0.92\linewidth]{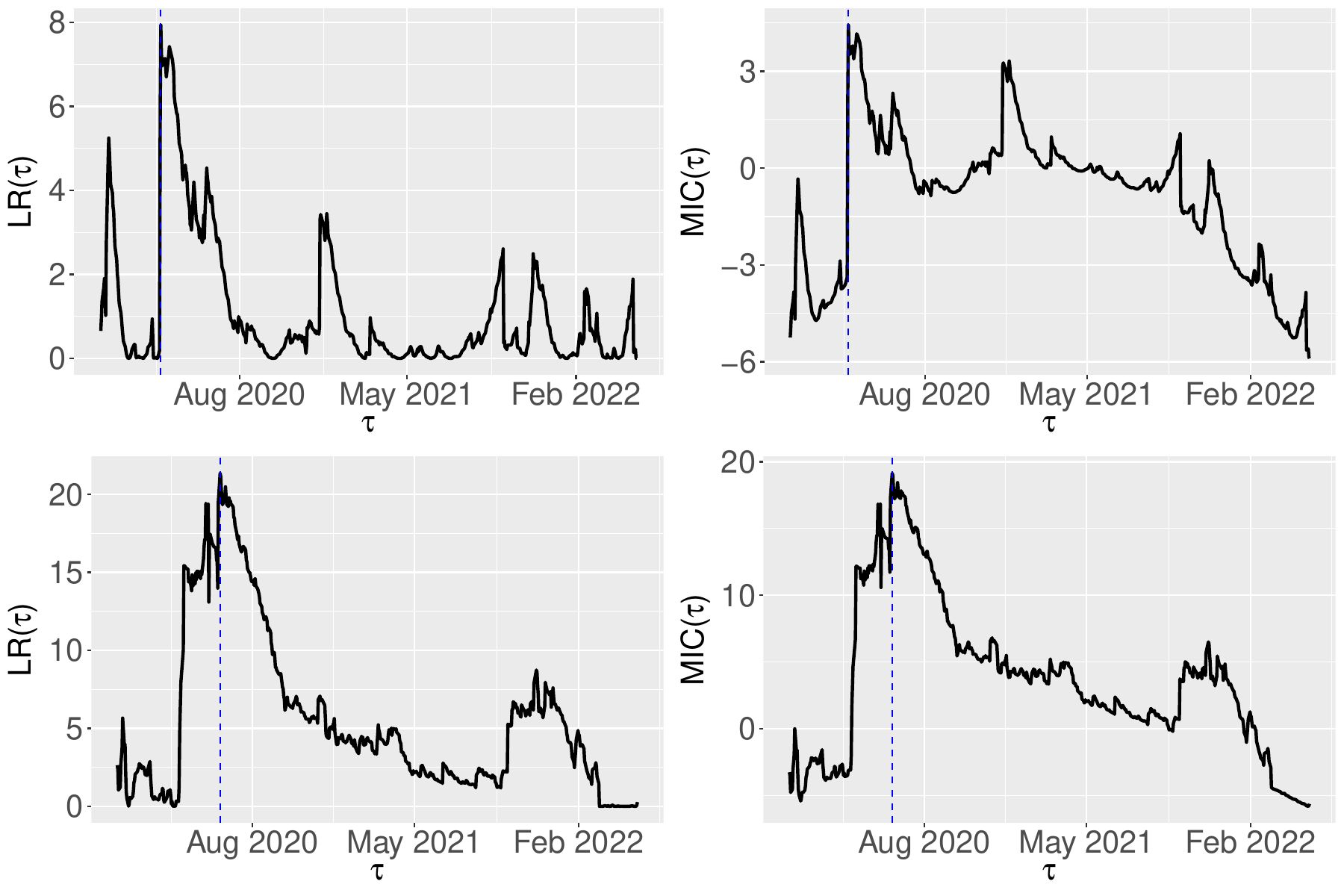}
    \caption{Temporal profiles of $\textrm{LR}(\tau)$ in \eqref{eq:loglik_diff} and $\textrm{MIC}(\tau)$ in \eqref{eq:mic_tau}, for the daily maximum (top) and minimum (bottom) rates of return series.}
    \label{fig:change_in_dependency}
\end{figure}

We first discuss the results for the daily maximum RoR series. In the case of LRT, we obtain a $p$-value of 0.027. Hence, we reject our null hypothesis $H_0$, i.e., no changepoint. We obtain the estimated changepoint to be $\widehat{\tau}^{\textrm{max}}_{\textrm{LRT}}=81$, which corresponds to the date March 26, 2020. Based on MIC, we obtain a $p$-value of 0.021. Hence, we also reject our null hypothesis $H_0$ based on MIC. We obtain the estimated changepoint to be $\widehat{\tau}^{\textrm{max}}_{\textrm{MIC}}=81$, which coincides with the result based on LRT. Top panels of Figure \ref{fig:change_in_dependency} present the profiles $\textrm{LR}(\tau)$ in \eqref{eq:loglik_diff} and $\textrm{MIC}(\tau)$ in \eqref{eq:mic_tau} for the daily maximum RoR series. Apart from $\widehat{\tau}^{\textrm{max}}_{\textrm{LRT}}=\widehat{\tau}^{\textrm{max}}_{\textrm{MIC}}=81$, certain other peaks are also observable in both profiles $\textrm{LR}(\tau)$ and $\textrm{MIC}(\tau)$; the most prominent peaks, where $\textrm{LR}(\tau)$ is higher than 6, occur between $\tau = 81$ and $\tau = 94$. Corresponding to $\tau=81$, the MLE of $\Lambda_1$ and $\Lambda_T$ are $\widehat{\Lambda}_1 = 1.012$ and $\widehat{\Lambda}_T = 1.540$, respectively. Ignoring the first and the last 50 days (due to the instability in estimation), for $\tau=81$, $\widehat{\Lambda}_T-\widehat{\Lambda}_1$ attains its highest value, which shows the correctness of the inference based on LRT and MIC.

We next discuss the results for the daily minimum RoR series. In the case of LRT, we obtain a $p$-value of less than 0.001. Hence, we reject our null hypothesis $H_0$, i.e., no changepoint. We obtain the estimated changepoint to be $\widehat{\tau}^{\textrm{min}}_{\textrm{LRT}}=128$, which corresponds to the date June 08, 2020. Based on MIC, we again obtain a $p$-value of less than 0.001. Hence, we also reject our null hypothesis $H_0$ based on MIC. We obtain the estimated changepoint to be $\widehat{\tau}^{\textrm{min}}_{\textrm{MIC}}=128$, which coincides with the result based on LRT. Bottom panels of Figure \ref{fig:change_in_dependency} present the profiles $\textrm{LR}(\tau)$ in \eqref{eq:loglik_diff} and $\textrm{MIC}(\tau)$ in \eqref{eq:mic_tau} for the daily minimum RoR series. Apart from $\widehat{\tau}^{\textrm{min}}_{\textrm{LRT}}=\widehat{\tau}^{\textrm{min}}_{\textrm{MIC}}=128$, certain other peaks are also observable in both profiles $\textrm{LR}(\tau)$ and $\textrm{MIC}(\tau)$; the most prominent peaks, where $\textrm{LR}(\tau)$ is higher than 18, occur between $\tau = 128$ and $\tau = 135$. Corresponding to $\tau=128$, the MLE of $\Lambda_1$ and $\Lambda_T$ are $\widehat{\Lambda}_1 = 0.682$ and $\widehat{\Lambda}_T = 1.003$, respectively.

According to \cite{bt2020Coronavirus}, the Government of India imposed a countrywide lockdown on March 25, 2020, which is very close to our estimated changepoint on March 26, 2020, for the daily maximum RoR series. This lockdown phase continued for 21 days, and all the high peaks in $\textrm{LR}(\tau)$ indicate this period. Before this date, we see less dependence (small $\Lambda_1$) between the daily maximum RoR of IndiGo and SpiceJet airlines due to the pre-lockdown situation. After the announcement of the lockdown, the aviation sector faced an unprecedented interruption, and both RoR behaved in a more similar fashion; this led to a higher $\Lambda_T$ than $\Lambda_1$. Despite the two companies focusing on different strategies to mitigate the challenges posed by COVID-19 as described in Section \ref{sec:Intro}, the effect of the initial phase of the lockdown appears to be the strongest one. On the other hand, the first phase of unlock started on June 1, 2020 and the estimated changepoint on June 8, 2020 for the daily minimum RoR series is also close to that date.

From a financial perspective, our daily maximum RoR analysis demonstrates that before the onset of the pandemic, investors used to distribute their investments in the two airlines with higher variability. After the onset of the pandemic, investors sold their stocks in both airlines, given the high vulnerability of the aviation industry in the pandemic situation, that can occur in future as well and the same investment philosophy for both airlines would persist. In the marginal scale, the RoR for IndiGo remained more stable than that for SpiceJet, but while looking at the dependence structure, both behave more similarly than before after the onset of the pandemic which affect all members of a vulnerable industry simultaneously.

\section{Discussions and Conclusions}
\label{sec:Conclusion}
The literature on changepoint estimation in the context of extreme value analysis is scarce. Some recent publications \citep{jaruvskova2008analysis, dierckx2010changepoint,nascimento2017bayesian, lattanzi2021change} focused on both frequentist and Bayesian perspectives of changepoint estimation for the univariate block maxima and threshold exceedances; however, as of our knowledge, no literature so far has focused on estimating changepoints in the extremal dependence structures or extreme value copulas except for a book chapter by \cite{da2004change}, where the authors focused on a specific copula, namely the bivariate Gumbel copula. Understanding the structural changes in the extremal dependence structure is often crucial for analyzing stock price data because the buy and sell positions generally occur when the prices reach high and low values, respectively. Share prices of two companies doing similar businesses are naturally dependent, specifically when both companies do business in a volatile sector like aviation, where a pandemic like COVID-19 can impact businesses severely. Given the bivariate H\"usler-Reiss distribution being the only possible limit of a sequence of bivariate Gaussian random variables, we explore different changepoint estimation strategies for this distribution. While the likelihood ratio test is the most popular approach in the literature, using simulation studies, we showcase that the hypothesis testing based on modified information criterion proposed by \cite{chen2006information} is generally more powerful. The most crucial changepoint in the daily maximum rate of return of the IndiGo and SpiceJet airlines, identified based on the methodology discussed here, almost coincides with the declaration of the first phase of the lockdown in India during COVID-19. This observation showcases the effectiveness of the two hypothesis testing procedures discussed here. During the announcement of the first phase of the lockdown, the number of COVID-19 cases was lower compared to the peaks of the three waves until mid-2022. Thus, the number of COVID-19 cases per day is not a meaningful predictor of the changepoint; rather, information on precautionary measures like lockdowns are significant factors while analyzing data related to the aviation industry.

In this paper, we focused on identifying the most significant changepoints for the upper and lower joint tails which would convey ideas on the most important factors behind changes during a period of turmoil like COVID-19. There is also a vast literature on estimating multiple changepoints \citep{barry1993bayesian, vexler2006guaranteed, killick2012optimal}. Among the algorithmic approaches, two most common approaches are binary segmentation \citep{barry1993bayesian} and Pruned Exact Linear Time \citep[PELT,][]{killick2012optimal}. Binary segmentation method recursively partitions the data into segments, detecting a single changepoint in each iteration, while PELT aims to find the minimum-cost partitioning of the data into segments using dynamic programming. Binary segmentation is very popular due to its easy implementation and may work fine in other applications like identification of changes in the mean structure, but require more observations for identifying any change in the dependence-related parameter of the bivariate H\"usler-Reiss distribution. In our case, before the first changepoint for the sequences of the daily maximum rate of return, there are very less number of observations and thus it is difficult to estimate the changepoint within this period with less uncertainty. Exploring PELT or other more efficient approaches like penalized regression methods or nonparametric methods would be a future research direction.

We explored the changepoint estimation focusing on the measures $\chi_U$ and $\chi_L$ in \eqref{eq:chi_u}. Apart from them, \cite{coles1999dependence} proposed related measures $\bar\chi_U$ and $\bar\chi_L$ given by
\begin{eqnarray}
    \nonumber \bar\chi_U(u) &=& 2\log\textrm{P}\{F_X(X)>u\}/\log\textrm{P}\{F_X(X)>u,F_Y(Y)>u\} - 1, \\
 \nonumber   \bar\chi_L(u) &=& 2\log\textrm{P}\{F_X(X)<u\}/\log\textrm{P}\{F_X(X)<u,F_Y(Y)<u\} - 1,~~~u \in (0,1),
\end{eqnarray}
where the notations correspond to \eqref{eq:chi_u}, and the limiting $\bar\chi$-measures are defined as $\bar\chi_U = \lim_{u \uparrow 1} \bar\chi_U(u)$ and $\bar\chi_L = \lim_{u \downarrow 0} \bar\chi_L(u)$. For the upper tail, there are two different extremal dependence regimes based on the two measures $\chi_U$ and $\bar{\chi}_U$; here $\chi_L$ and $\bar{\chi}_L$ similarly defines two regimes in the lower tail. In the first regime, two random variables are called extremally dependent and here $\chi_U \in (0,1)$ and $\bar{\chi}_U = 1$. In the second regime, two random variables are called extremally independent and here $\chi_U = 0$ and $\bar{\chi}_U \in [-1,1)$. For the bivariate H\"usler-Reiss distribution, except for the trivial case of independence, the first class holds and as a result, $\bar{\chi}_U=\bar{\chi}_L=1$ for both before and after the changepoint. Hence, a method of changepoint detection based on $\bar{\chi}_U$ or $\bar{\chi}_L$ cannot be used under the modeling assumption in this paper, i.e., the class of bivariate H\"usler-Reiss distribution.

While we analyzed data for two specific airlines IndiGo and SpiceJet that acquired the highest market shares in the Indian aviation industry during COVID-19, our methodology can be adapted easily for analyzing data from other airlines, other countries, and other business sectors. Apart from a pandemic, one can also use our methodology in case of a recession. From a methodological perspective, one can adapt our approach to a general multivariate extreme value analysis beyond the bivariate case and also for analyzing spatial extremes, using pairwise likelihood \citep{huser2013composite}, where each component is a bivariate H\"usler-Reiss density. Further, graphical approaches for multivariate \citep{engelke2020graphical} and spatial extremes \citep{cisneros2023spatial} use the bivariate H\"usler-Reiss densities for each edge of a graph. Extending our methodology for detecting changepoints in a sequence of multivariate or spatial extremes would be a future endeavor. Any theoretical proposition corresponding to consistency with respect to the bivariate H\"sler-Reiss distribution has not been explored in this paper, except for discussing it numerically only in Section \ref{subsec:consistency}. Exploring such properties theoretically would be an interesting future endeavor.

\section*{Data availability statement}
The dataset analyzed in this article is available at \url{https://www.investing.com/}.

\section*{Disclosure statement}
No potential conflict of interest was reported by the authors.

\bibliographystyle{tfs}
\bibliography{references}

\begin{thebibliography}{10}
\providecommand{\MR}{\relax\unskip\space MR }
\providecommand{\url}[1]{\normalfont{#1}}
\providecommand{\urlprefix}{Available at }

\bibitem{agrawal2021sustainability}
A. Agrawal, \emph{Sustainability of airlines in {I}ndia with {COVID}-19: Challenges ahead and possible way-outs}, Journal of Revenue and Pricing Management 20 (2021), pp. 1--16.

\bibitem{banerji2016case}
D. Banerji, P. Mukherjee, and N. Siroya, \emph{Case study-soaring into the high skies}, SSRN 2873325  (2016).

\bibitem{barry1993bayesian}
D. Barry and J.A. Hartigan, \emph{A bayesian analysis for change point problems}, Journal of the American Statistical Association 88 (1993), pp. 309--319.

\bibitem{bosc2003automatic}
M. Bosc, F. Heitz, J.P. Armspach, I. Namer, D. Gounot, and L. Rumbach, \emph{Automatic change detection in multimodal serial {MRI}: application to multiple sclerosis lesion evolution}, NeuroImage 20 (2003), pp. 643--656.

\bibitem{brown1977extreme}
B.M. Brown and S.I. Resnick, \emph{Extreme values of independent stochastic processes}, Journal of Applied Probability 14 (1977), pp. 732--739.

\bibitem{bt2020indigo}
 BT, \emph{Coronavirus impact: How {IndiGo} is turning {COVID} crisis into an opportunity}, Business Today  (2020). \urlprefix\url{https://www.businesstoday.in/amp/bt-buzz/news/story/coronavirus-impact-how-indigo-is-turning-covid-crisis-into-an-opportunity-260283-2020-06-05"}.

\bibitem{cai2016changepoint}
X. Cai, K.K. Said, and W. Ning, \emph{Changepoint analysis with bathtub shape for the exponential distribution}, Journal of Applied Statistics 43 (2016), pp. 2740--2750. \urlprefix\url{https://doi.org/10.1080/02664763.2016.1143455}.

\bibitem{campbell1973asymptotic}
J.W. Campbell and C.P. Tsokos, \emph{The asymptotic distribution of maxima in bivariate samples}, Journal of the American Statistical Association 68 (1973), pp. 734--739. \urlprefix\url{http://www.jstor.org/stable/2284810}.

\bibitem{cartea2014buy}
{\'A}. Cartea, S. Jaimungal, and J. Ricci, \emph{Buy low, sell high: A high frequency trading perspective}, SIAM Journal on Financial Mathematics 5 (2014), pp. 415--444.

\bibitem{valerie2016extreme}
V. Chavez-Demoulin, P. Embrechts, and M. Hofert, \emph{An extreme value approach for modeling operational risk losses depending on covariates}, The Journal of Risk and Insurance 83 (2016), pp. 735--776. \urlprefix\url{http://www.jstor.org/stable/43998282}.

\bibitem{chen2006information}
J. Chen, A. Gupta, and J. Pan, \emph{Information criterion and change point problem for regular models}, Sankhyā: The Indian Journal of Statistics (2003-2007) 68 (2006).

\bibitem{chen1997testing}
J. Chen and A.K. Gupta, \emph{Testing and locating variance changepoints with application to stock prices}, Journal of the American Statistical Association 92 (1997), pp. 739--747. \urlprefix\url{http://www.jstor.org/stable/2965722}.

\bibitem{chen2012parametric}
J. Chen and A. Gupta, \emph{Parametric Statistical Change Point Analysis: With Applications to Genetics, Medicine, and Finance}, Springer, 2012.

\bibitem{chernoff1964estimating}
H. Chernoff and S. Zacks, \emph{Estimating the current mean of a normal distribution which is subjected to changes in time}, The Annals of Mathematical Statistics 35 (1964), pp. 999 -- 1018. \urlprefix\url{https://doi.org/10.1214/aoms/1177700517}.

\bibitem{cherubini2004copula}
U. Cherubini, E. Luciano, and W. Vecchiato, \emph{Copula methods in finance}, John Wiley \& Sons, 2004.

\bibitem{cisneros2023spatial}
D. Cisneros, A. Hazra, and R. Huser, \emph{Spatial wildfire risk modeling using mixtures of tree-based multivariate {P}areto distributions}, arXiv preprint arXiv:2308.03870  (2023).

\bibitem{coles2001introduction}
S. Coles, \emph{An introduction to statistical modeling of extreme vaues}, Springer Series in Statistics, Springer, New York  (2001), pp. XIV, 209. \urlprefix\url{https://doi.org/10.1007/978-1-4471-3675-0}.

\bibitem{coles1994statistical}
S.G. Coles and J.A. Tawn, \emph{Statistical methods for multivariate extremes: An application to structural design}, Journal of the Royal Statistical Society. Series C (Applied Statistics) 43 (1994), pp. 1--48. \urlprefix\url{http://www.jstor.org/stable/2986112}.

\bibitem{cooley2006variograms}
D. Cooley, P. Naveau, and P. Poncet, \emph{Variograms for spatial max-stable random fields}, in \emph{Dependence in probability and statistics}, Springer, New York, NY,  2006, pp. 373--390.

\bibitem{csorgo1997limit}
M. Cs{\"o}rg{\"o} and L. Horv{\'a}th, \emph{Limit theorems in change-point analysis}, Wiley series in probability and statistics, John Wiley and Sons, Chichester, 1997.

\bibitem{david2004order}
H.A. David and H.N. Nagaraja, \emph{Order statistics}, John Wiley and Sons, New Jersey, 2004.

\bibitem{davison2015statistics}
A.C. Davison and R. Huser, \emph{Statistics of extremes}, Annual Review of Statistics and its Application 2 (2015), pp. 203--235.

\bibitem{davison2013geostatistics}
A.C. Davison, R. Huser, and E. Thibaud, \emph{Geostatistics of dependent and asymptotically independent extremes}, Mathematical Geosciences 45 (2013), pp. 511--529.

\bibitem{de2020tracking}
M. de  Carvalho, M. Leonelli, and A. Rossi, \emph{Tracking change-points in multivariate extremes}, arXiv preprint arXiv:2011.05067  (2020).

\bibitem{da2004change}
A.D.C. Dias and P. Embrechts, \emph{Change-point analysis for dependence structures in finance and insurance}, in \emph{Risk Measures for the 21st Century}, G.P. Szeg{\"o}, ed., chap.~16, Wiley, Chichester,  2004, pp. 321--335.

\bibitem{dierckx2010changepoint}
G. Dierckx and J.L. Teugels, \emph{Changepoint analysis of extreme values}, Environmetrics 21 (2010), pp. 661--686.

\bibitem{nascimento2017bayesian}
F.F. do  Nascimento and W.V.M. e~ Silva, \emph{A {Bayesian} model for multiple change point to extremes, with application to environmental and financial data}, Journal of Applied Statistics 44 (2017), pp. 2410--2426.

\bibitem{eiauer1978use}
P. Eiauer and P. Hackl, \emph{The use of mosums for quality control}, Technometrics 20 (1978), pp. 431--436.

\bibitem{embrechts1997modelling}
P. Embrechts, C. Kl{\"u}ppelberg, and T. Mikosch, \emph{Modelling extremal events: for insurance and finance}, Springer Science and Business Media, Berlin, 1997.

\bibitem{engelke2020graphical}
S. Engelke and A.S. Hitz, \emph{Graphical models for extremes}, Journal of the Royal Statistical Society Series B: Statistical Methodology 82 (2020), pp. 871--932.

\bibitem{engelke2021sparse}
S. Engelke and J. Ivanovs, \emph{Sparse structures for multivariate extremes}, Annual Review of Statistics and Its Application 8 (2021), pp. 241--270.

\bibitem{fisher1928limiting}
R.A. Fisher and L.H.C. Tippett, \emph{Limiting forms of the frequency distribution of the largest or smallest member of a sample}, Mathematical Proceedings of the Cambridge Philosophical Society 24 (1928), pp. 180--190.

\bibitem{gardner1969detecting}
L.A. Gardner, \emph{On detecting changes in the mean of normal variates}, The Annals of Mathematical Statistics 40 (1969), pp. 116 -- 126. \urlprefix\url{https://doi.org/10.1214/aoms/1177697808}.

\bibitem{gong2022asymmetric}
Y. Gong and R. Huser, \emph{Asymmetric tail dependence modeling, with application to cryptocurrency market data}, The Annals of Applied Statistics 16 (2022), pp. 1822--1847.

\bibitem{gormsen2020coronavirus}
N.J. Gormsen and R.S. Koijen, \emph{Coronavirus: Impact on stock prices and growth expectations}, The Review of Asset Pricing Studies 10 (2020), pp. 574--597.

\bibitem{gurevich2005change}
G. Gurevich and A. Vexler, \emph{Change point problems in the model of logistic regression}, Journal of Statistical Planning and Inference 131 (2005), pp. 313--331.

\bibitem{gurevich2010retrospective}
G. Gurevich and A. Vexler, \emph{Retrospective change point detection: from parametric to distribution free policies}, Communications in Statistics—Simulation and Computation 39 (2010), pp. 899--920.

\bibitem{hasan2014information}
A. Hasan, W. Ning, and A. Gupta, \emph{An information-based approach to the change-point problem of the noncentral skew-t distribution with applications to stock market data}, Sequential Analysis 33 (2014), pp. 458--474.

\bibitem{hashorva2010residual}
E. Hashorva, \emph{On the residual dependence index of elliptical distributions}, Statistics \& Probability Letters 80 (2010), pp. 1070--1078.

\bibitem{hawkins1992detecting}
D. Hawkins, \emph{Detecting shifts in functions of multivariate location and covariance parameters}, Journal of Statistical Planning and Inference 33 (1992), pp. 233--244. \urlprefix\url{https://www.sciencedirect.com/science/article/pii/0378375892900709}.

\bibitem{hosking1985estimation}
J.R.M. Hosking, J.R. Wallis, and E.F. Wood, \emph{Estimation of the generalized extreme-value distribution by the method of probability-weighted moments}, Technometrics 27 (1985), pp. 251--261.

\bibitem{hsu1977tests}
D. Hsu, \emph{Tests for variance shift at an unknown time point}, Journal of The Royal Statistical Society Series C-Applied Statistics 26 (1977), pp. 279--284.

\bibitem{huang2020lifetable}
F. Huang, R. Maller, and X. Ning, \emph{Modelling life tables with advanced ages: An extreme value theory approach}, Insurance: Mathematics and Economics 93 (2020), pp. 95--115. \urlprefix\url{https://www.sciencedirect.com/science/article/pii/S0167668720300482}.

\bibitem{huser2013composite}
R. Huser and A.C. Davison, \emph{Composite likelihood estimation for the {Brown--Resnick} process}, Biometrika 100 (2013), pp. 511--518.

\bibitem{Husler1989maxima}
J. H\"usler and R.D. Reiss, \emph{Maxima of normal random vectors: Between independence and complete dependence}, Statistics and Probability Letters 7 (1989), pp. 283--286.

\bibitem{icao2020effects}
 ICAO, \emph{{Effects of Novel Coronavirus ({COVID}-19) on Civil Aviation: Economic Impact Analysis}}, International Civil Aviation Organization Report  (2020). \urlprefix\url{https://www.icao.int/sustainability/Documents/COVID-19/ICAO_Coronavirus_Econ_Impact.pdf"}.

\bibitem{inclan1993detection}
C. Incl\'an, \emph{Detection of multiple changes of variance using posterior odds}, Journal of Business and Economic Statistics 11 (1993), pp. 289--300.

\bibitem{bt2020Coronavirus}
M. Jaiswal, \emph{{Coronavirus in India: 21-day lockdown begins; key highlights of PM Modi's speech}}, Business Today  (2020). \urlprefix\url{https://www.businesstoday.in/latest/economy-politics/story/coronavirus-in-india-21-day-lockdown-begins-key-highlights-of-pm-modi-speech-253038-2020-03-25}.

\bibitem{jaruvskova2008analysis}
D. Jaru{\v{s}}kov{\'a} and M. Rencov{\'a}, \emph{Analysis of annual maximal and minimal temperatures for some {European} cities by change point methods}, Environmetrics 19 (2008), pp. 221--233.

\bibitem{jaworski2010copula}
P. Jaworski, F. Durante, W.K. Hardle, and T. Rychlik, \emph{Copula theory and its applications}, Vol. 198, Springer, 2010.

\bibitem{kabluchko2009stationary}
Z. Kabluchko, M. Schlather, and L. De~Haan, \emph{Stationary max-stable fields associated to negative definite functions}, Ann. Probab. 37 (2009), pp. 2042–--2065.

\bibitem{killick2014changepoint}
R. Killick and I. Eckley, \emph{changepoint: {A}n {R} package for changepoint analysis}, Journal of Statistical Software 58 (2014), pp. 1--19.

\bibitem{killick2012optimal}
R. Killick, P. Fearnhead, and I.A. Eckley, \emph{Optimal detection of changepoints with a linear computational cost}, Journal of the American Statistical Association 107 (2012), pp. 1590--1598.

\bibitem{lai1995sequential}
T.L. Lai, \emph{Sequential changepoint detection in quality control and dynamical systems}, Journal of the Royal Statistical Society: Series B (Methodological) 57 (1995), pp. 613--644.

\bibitem{lattanzi2021change}
C. Lattanzi and M. Leonelli, \emph{A change-point approach for the identification of financial extreme regimes}, Brazilian Journal of Probability and Statistics 35 (2021), pp. 811--837.

\bibitem{lavielle2005using}
M. Lavielle, \emph{Using penalized contrasts for the change-point problem}, Signal Processing 85 (2005), pp. 1501--1510.

\bibitem{leadbetter1983extremes}
M.R. Leadbetter, G. Lindgren, and H. Rootz{\'e}n, \emph{Extremes and related properties of random sequences and processes}, Springer Science and Business Media, New York, 1983.

\bibitem{leadbetter1988extremal}
M. Leadbetter and H. Rootzen, \emph{Extremal theory for stochastic processes}, The Annals of Probability  (1988), pp. 431--478.

\bibitem{lerman1989detecting}
Z. Lerman and E. Schechtman, \emph{Detecting a change in the correlation coefficient in a sequence of bivariate normal variables}, Communications in Statistics - Simulation and Computation 18 (1989), pp. 589--599. \urlprefix\url{https://doi.org/10.1080/03610918908812778}.

\bibitem{liu2009changepoint}
Z. Liu and L. Qian, \emph{Changepoint estimation in a segmented linear regression via empirical likelihood}, Communications in Statistics - Simulation and Computation 39 (2009), pp. 85--100.

\bibitem{molina2008application}
M. Molina-Garcia, A. Fernandez-Duran, and J.I. Alonso, \emph{Application of extreme value distribution to model propagation fading in indoor mobile radio environments}, in \emph{2008 IEEE Radio and Wireless Symposium}. IEEE, 2008, pp. 97--100.

\bibitem{muggeo2011efficient}
V.M. Muggeo and G. Adelfio, \emph{Efficient change point detection for genomic sequences of continuous measurements}, Bioinformatics 27 (2011), pp. 161--166.

\bibitem{nelsen2006introduction}
R.B. Nelsen, \emph{An introduction to copulas}, Springer, 2006.

\bibitem{ngunkeng2014information}
G. Ngunkeng and W. Ning, \emph{Information approach for the change-point detection in the skew normal distribution and its applications}, Sequential Analysis 33 (2014), pp. 475--490. \urlprefix\url{https://doi.org/10.1080/07474946.2014.961845}.

\bibitem{owen1988empirical}
A.B. Owen, \emph{Empirical likelihood ratio confidence intervals for a single functional}, Biometrika 75 (1988), pp. 237--249.

\bibitem{page1955test}
E.S. Page, \emph{A test for a change in a parameter occurring at an unknown point}, Biometrika 42 (1955), pp. 523--527. \urlprefix\url{http://www.jstor.org/stable/2333401}.

\bibitem{page1954continuous}
E.S. Page, \emph{Continuous inspection schemes}, Biometrika 41 (1954), pp. 100--115.

\bibitem{patton2012review}
A.J. Patton, \emph{A review of copula models for economic time series}, Journal of Multivariate Analysis 110 (2012), pp. 4--18.

\bibitem{asoka2003epidemic}
A. Ramanayake and A.K. Gupta, \emph{Tests for an epidemic change in a sequence of exponentially distributed random variables}, Biometrical Journal 45 (2003), pp. 946--958. \urlprefix\url{https://onlinelibrary.wiley.com/doi/abs/10.1002/bimj.200390062}.

\bibitem{reeves2007review}
J. Reeves, J. Chen, X.L. Wang, R. Lund, and Q.Q. Lu, \emph{A review and comparison of changepoint detection techniques for climate data}, Journal of Applied Meteorology and Climatology 46 (2007), pp. 900--915.

\bibitem{rocco2014extreme}
M. Rocco, \emph{Extreme value theory in finance: A survey}, Journal of Economic Surveys 28 (2014), pp. 82--108. \urlprefix\url{https://onlinelibrary.wiley.com/doi/abs/10.1111/j.1467-6419.2012.00744.x}.

\bibitem{said2017likelihood}
K.K. Said, W. Ning, and Y. Tian, \emph{Likelihood procedure for testing changes in skew normal model with applications to stock returns}, Communications in Statistics - Simulation and Computation 46 (2017), pp. 6790--6802. \urlprefix\url{https://doi.org/10.1080/03610918.2016.1212067}.

\bibitem{said2019modified}
K.K. Said, W. Ning, and Y. Tian, \emph{{Modified information criterion for testing changes in skew normal model}}, Brazilian Journal of Probability and Statistics 33 (2019), pp. 280 -- 300. \urlprefix\url{https://doi.org/10.1214/17-BJPS388}.

\bibitem{sibuya1960bivariate}
M. Sibuya, \emph{Bivariate extreme statistics}, Annals of the Institute of Statistical Mathematics 11 (1960), pp. 195--210. \urlprefix\url{https://doi.org/10.1007/BF01682329}.

\bibitem{kaur2021impact}
P.K. Sidhu and R. Shukla, \emph{{Impact of the COVID-19 pandemic on the Indian domestic aviation industry}}, in \emph{2021 Reconciling Data Analytics, Automation, Privacy, and Security: A Big Data Challenge (RDAAPS)}. IEEE, 2021, pp. 1--8.

\bibitem{srivastava1986likelihood}
M. Srivastava and K.J. Worsley, \emph{Likelihood ratio tests for a change in the multivariate normal mean}, Journal of the American Statistical Association 81 (1986), pp. 199--204.

\bibitem{stephenson2002evd}
A.G. Stephenson, \emph{evd: Extreme value distributions}, R news 2 (2002), pp. 31--32.

\bibitem{coles1999dependence}
J.H. Stuart~Coles and J. Tawn, \emph{Dependence measures for extreme value analyses}, Extremes 2 (1999), pp. 339--365. \urlprefix\url{https://doi.org/10.1023/A:1009963131610}.

\bibitem{tawn1988bivariate}
J.A. Tawn, \emph{Bivariate extreme value theory: Models and estimation}, Biometrika 75 (1988), pp. 397--415. \urlprefix\url{http://www.jstor.org/stable/2336591}.

\bibitem{thies2018bayesian}
S. Thies and P. Moln{\'a}r, \emph{Bayesian change point analysis of {Bitcoin} returns}, Finance Research Letters 27 (2018), pp. 223--227.

\bibitem{oliveira1989statistical}
J. Tiago~de  Oliveira, \emph{Statistical decision for bivariate extremes}, in \emph{Extreme Value Theory: Proceedings of a Conference held in Oberwolfach, Dec. 6--12, 1987}. Springer, 1989, pp. 246--261.

\bibitem{tian2023changepoint}
W. Tian, L. Pang, C. Tian, and W. Ning, \emph{Changepoint analysis for {Kumaraswamy} distribution}, Mathematics 11 (2023). \urlprefix\url{https://www.mdpi.com/2227-7390/11/3/553}.

\bibitem{tian2022changepoint}
W. Tian and Y. Yang, \emph{Changepoint analysis for weighted exponential distribution}, Communications in Statistics - Simulation and Computation 0 (2022), pp. 1--13. \urlprefix\url{https://doi.org/10.1080/03610918.2021.2020288}.

\bibitem{tibshirani1987local}
R. Tibshirani and T. Hastie, \emph{Local likelihood estimation}, Journal of the American Statistical Association 82 (1987), pp. 559--567.

\bibitem{toi2020indigo}
 TOI, \emph{Lockdown impact: {IndiGo} reports {INR} 2,844 crore loss in {A}pril}, Times of India  (2020). \urlprefix\url{https://timesofindia.indiatimes.com/business/india-business/lockdown-impact-indigo-reports-rs-2844-crore-loss-in-april-june-quarter/articleshow/77240430.cms}.

\bibitem{vexler2006guaranteed}
A. Vexler, \emph{Guaranteed testing for epidemic changes of a linear regression model}, Journal of Statistical Planning and Inference 136 (2006), pp. 3101--3120.

\bibitem{vexler2009optimal}
A. Vexler and C. Wu, \emph{An optimal retrospective change point detection policy}, Scandinavian Journal of Statistics 36 (2009), pp. 542--558.

\bibitem{worsley1979likelihood}
K.J. Worsley, \emph{On the likelihood ratio test for a shift in location of normal populations}, Journal of the American Statistical Association 74 (1979), pp. 365--367.

\bibitem{worsley1983power}
K. Worsley, \emph{The power of likelihood ratio and cumulative sum tests for a change in a binomial probability}, Biometrika 70 (1983), pp. 455--464.

\bibitem{zhao2013changepoint}
H. Zhao, H. Chen, and W. Ning, \emph{Changepoint analysis by modified empirical likelihood method in two-phase linear regression models}, Open Journal of Applied Sciences 03 (2013), pp. 1--6.

\bibitem{zou2007empirical}
C. Zou, Y. Liu, P. Qin, and Z. Wang, \emph{Empirical likelihood ratio test for the change-point problem}, Statistics and Probability Letters 77 (2007), pp. 374--382.

\bibitem{zwiers1998changes}
F.W. Zwiers and V.V. Kharin, \emph{Changes in the extremes of the climate simulated by {CCC GCM2} under {CO2} doubling}, Journal of Climate 11 (1998), pp. 2200 -- 2222. \urlprefix\url{https://journals.ametsoc.org/view/journals/clim/11/9/1520-0442_1998_011_2200_citeot_2.0.co_2.xml}.

\end{thebibliography}

\end{document}